\documentstyle[12pt,aaspp4]{article}

\slugcomment{To appear in {\it The Astrophysical Journal}}

\lefthead{Garc\'\i a L\'opez et al.}
\righthead{Boron in very metal-poor stars}

\begin{document}

\title{Boron in Very Metal-Poor Stars\footnote{Based on 
observations obtained with the NASA/ESA Hubble Space Telescope (HST) at the
Space Telescope Science Institute, which is operated by the Association of
Universities for Research in Astronomy, Inc., under NASA contract
NAS5-26555}}

\author{Ram\'on J. Garc\'\i a L\'opez\altaffilmark{2}, David L. 
Lambert\altaffilmark{3}, Bengt Edvardsson\altaffilmark{4}, Bengt
Gustafsson\altaffilmark{4}, Dan Kiselman\altaffilmark{5}, and Rafael
Rebolo\altaffilmark{2}}

\authoraddr{Instituto de Astrof\'\i sica de Canarias, E-38200 La Laguna,
Tenerife, Spain}

\authoremail{rgl@ll.iac.es, dll@astro.as.utexas.edu, 
bengt.edvardsson@astro.uu.se, bengt.gustafsson@astro.uu.se, 
dan@surge.astro.su.se, rrl@ll.iac.es}

\altaffiltext{2}{Instituto de Astrof\'\i sica de Canarias, E-38200 La Laguna,
Tenerife, Spain}
\altaffiltext{3}{Department of Astronomy, University of Texas, RLM 15.308,
Austin, TX 78712, USA}
\altaffiltext{4}{Uppsala Astronomical Observatory, Box 515, SE-751 20 Uppsala,
Sweden}
\altaffiltext{5}{The Royal Academy of Sciences, Stockholm Observatory, SE-133 
36 Saltsj\"obaden, Sweden}

\newpage

\begin{abstract}

We have observed the \ion{B}{1} $\lambda$ 2497 \AA\ line to derive the boron
abundances of two very metal-poor stars selected to help in tracing the origin
and evolution of this element in the early Galaxy: BD $+$23\arcdeg 3130 and HD
84937. The observations were conducted using the Goddard High Resolution
Spectrograph on board the {\it Hubble Space Telescope}. A very detailed
abundance analysis via spectral synthesis has been carried out for these two
stars, as well as for two other metal-poor objects with published spectra,
using both Kurucz and OSMARCS model photospheres, and taking into account
consistently the NLTE effects on the line formation. We have also re-assessed
all published boron abundances of old disk and halo unevolved stars. Our
analysis shows that the combination of high effective temperature ($T_{\rm
eff}\gtrsim 6000$ K, for which boron is mainly ionized) and low metallicity
([Fe/H] $\lesssim -1$) makes it difficult to obtain accurate estimates of boron
abundances from the \ion{B}{1} $\lambda$ 2497 \AA\ line. This is the case of HD
84937 and three other published objects (including two stars with [Fe/H] $\sim
-3$), for which only upper limits can be established. BD $+$23\arcdeg 3130,
with [Fe/H] $\sim -2.9$ and log N(B)$_{\rm NLTE}=0.05 \pm 0.30$, appears then
as the most metal-poor star for which a firm measurement of the boron abundance
presently exists. The evolution of the boron abundance with metallicity that
emerges from the seven remaining stars with $T_{\rm eff}<6000$ K and [Fe/H]
$<-1$, for which beryllium abundances were derived using the same stellar
parameters, shows a linear increase with a slope $\sim 1$. Furthermore, the
B/Be ratio found is constant at a value $\sim 20$ for stars in the range
$-3<[{\rm Fe/H}]<-1$. These results point to spallation reactions of ambient
protons and $\alpha$ particles with energetic particles enriched in CNO as the
origin of boron and beryllium in halo stars. 

\end{abstract}
 
\keywords{Galaxy: evolution --- nuclear reactions, nucleosynthesis, abundances
--- stars: abundances --- stars: individual (BD $+23$\arcdeg 3130, HD 84937)
--- stars: late-type --- stars: Population II}

\newpage

\section{Introduction} 
\label{sec1}

Understanding the origins of the chemical elements is a fundamental goal of
contemporary astrophysics. As this understanding has advanced in  recent years,
a few elements have been rather neglected. Boron is one such element. A primary
reason for its neglect has been that stellar absorption lines of the boron atom
and ions are not detectable in spectra obtainable from the ground;  the  low
abundance of boron means that this element is detectable only through its
strong resonance lines which are in the ultraviolet. The advance in ultraviolet
stellar spectroscopy made possible by the {\it Hubble Space Telescope} (HST)
and its Goddard High Resolution Spectrograph (GHRS) has led to new data on the
boron abundance of cool and hot stars.

Pioneering work on stellar boron abundances was presented by Boesgaard \&
Heacox (1978) who analyzed the \ion{B} {2} $\lambda$ 1362 \AA\ resonance line
in spectra of early-type stars observed with the {\it Copernicus} satellite. To
map the evolution of boron with metallicity it is necessary to observe
unevolved metal-poor stars. Since these are cool, boron detection has to be
made using the atomic resonance lines. No advance in our knowledge of the boron
abundance came with the {\it International Ultraviolet Explorer} (IUE)
satellite (an attempt to detect the \ion{B}{1} resonance lines near $\lambda$
2497 \AA\ in a metal-poor star was made by Molaro 1987). Duncan, Lambert, \&
Lemke's (1992) detection and analysis of the \ion{B}{1} $\lambda$ 2497 \AA\
lines in three metal-poor stars observed with the GHRS has been followed by
other studies of these lines -- see Edvardsson et al. (1994), Rebull et al.
(1996), and Duncan et al. (1997a, 1998). Kiselman (1994) and Kiselman \&
Carlsson (1996) showed that non-local thermodynamic equilibrium (NLTE) effects
have to be considered in analyses of the 2497 \AA\ lines.

Standard big bang nucleosynthesis calculations do not predict a significant
production of $^6$Li, $^9$Be, $^{10}$B, and $^{11}$B, which are thought to be
produced by spallation reactions between cosmic rays and CNO nuclei (see
Ramaty, Kozlovsky, \& Lingenfelter 1996 and Ramaty et al. 1997 for recent
discussions on the theoretical yields). Models of inhomogeneous big bang
nucleosynthesis suggest that observable amounts of Be and B could be produced
(e.g. Malaney \& Fowler 1989; Kajino \& Boyd 1990; Jedamzik et al. 1994), but
this possibility has been questioned on the theoretical grounds (e.g. Terasawa
\& Sato 1990; Reeves et al. 1990).  

In this paper, we derive the boron abundances of two metal-poor stars observed
by us to fill in specific places of the study of the evolution of boron. BD
$+23$\arcdeg 3130 is a cool very metal-poor star ([Fe/H]$\sim -2.9$) selected
to extend B abundance determinations to a star of a factor of two more
deficient than HD 140283 ([Fe/H]$\sim -2.7$; previously studied by Duncan et al.
1992 and Edvardsson et al. 1994), and is probably the most metal-poor slightly
evolved star for which boron can be measured with HST. Duncan et al. (1997a)
has also observed two stars (BD $-13$\arcdeg 3442 and BD $+$3\arcdeg 740) with
metallicities [Fe/H]$\sim -3$; however, they are much warmer and, hence, boron
is mainly ionized making more difficult to obtain a precise measurement of its
abundance from the \ion{B}{1} lines. Our second star, HD 84937, is of special
interest because $^6$Li (another expected product of spallation and fusion
reactions driven by cosmic rays) has been traced in its atmosphere (Smith,
Lambert, \& Nissen 1993; Hobbs \& Thorburn 1994, 1997). Both stars also have
previous measurements of $^7$Li in the literature, and have been included in
the samples for deriving beryllium and oxygen (from UV OH lines) abundances in
metal-poor stars by Garc\'\i a L\'opez et al. (1998) and Israelian, Garc\'\i a
L\'opez, \& Rebolo (1998), respectively, using the same stellar parameters.

To better study the dependence of the derived abundances on the tools usually
employed in the analyses, we present here the first detailed comparison of
boron abundances derived using the two sets of model photospheres widely
employed in the analyses of stellar spectra: those calculated with the ATLAS
program, constructed by R. L. Kurucz, and those from the MARCS program,
developed at Uppsala Astronomical Observatory. In both cases there have been
considerable recent developments in these codes and their underlying input
data. This careful analysis of boron abundances in very metal-poor stars, in
which we also re-assess published abundances of old disk and halo unevolved
stars, is aimed at better constraining observationally the nucleosynthetic
origins of boron.

\section{Observations}
\label{sec2}

Observations of the stars BD $+23$\arcdeg 3130 and HD 84937 were carried out
using the GHRS  in the low resolution G270M/SSA mode ($\lambda
/\Delta\lambda\sim 25000$). The G270M grating was centered at $\lambda$ 2497
\AA\ to observe the \ion{B}{1} doublet located at $\lambda\lambda$ 2496.772 and
2497.723 \AA\ (air values). The standard FP-SPLIT mode, which provides spectra
in slightly different carrousel positions, was used to obtain high
signal-to-noise ratios (S/N) and minimize the effects of the photocathode
glitches and the geomagnetically-induced image motion problem (GIMP). Exposure
times were estimated from the UV fluxes measured by IUE at $\lambda$ 2500 \AA\
for our targets. BD $+23$\arcdeg 3130 is a cool metal-poor giant with $B= 9.59$
and was observed with HST during ten orbits (21434 sec exposure time) on 1995
August 4, while HD 84937 is a hotter metal-poor dwarf with $B=8.69$ which
required three orbits (5440 sec) on 1996 March 21.

Data reductions were performed by standard procedures using the ``stsdas''
package of the IRAF\footnote{IRAF is distributed by National Optical Astronomy
Observatories, which is operated by the Association of Universities for
Research in Astronomy, Inc., under contract with the National Science
Foundation, USA.} suite of programs. Final S/N values of 100 and 125 per diode 
(50 and 63 per pixel) were achieved for BD $+23$\arcdeg 3130 and HD 84937,
respectively. Their corresponding measured continuum fluxes at $\lambda$ 2500
\AA\ were $2.56\times 10^{-13}$ and $1.46\times 10^{-12}$ erg cm$^{-2}$
s$^{-1}$ \AA$^{-1}$. These values are about 30--40\% larger than those measured
by IUE at that wavelength, and these differences are reduced to 20--25\% when
the HST data are degraded to match the IUE spectral resolution. Differences in
the flux calibrations of both instruments should explain these discrepancies.

The wavelength calibration provided a dispersion of 0.024 \AA\ pixel$^{-1}$,
and the final spectra were shifted in wavelength to correct for stellar and 
Earth motions using the \ion{Fe}{1} $\lambda$ 2496.533 \AA\ line as the 
reference. Continuum normalization is a difficult task in a crowded spectral 
region like this one. In principle, the HST spectra have been corrected for
the spectrograph efficiency versus wavelength. A soft curve was fitted to the
upper points of the spectra, assigning a value of 1 to those pixels
corresponding to the region just shortwards of $\lambda$ 2500 \AA, considered
by Duncan et al. (1992; 1997b) and Edvardsson et al. (1994) as the most
line-free region in the data. Spectral syntheses performed afterwards showed
that this criterion agrees well with the predictions of synthetic spectra.
Figure 1 shows a portion of the final spectra for both stars.


\section{Stellar Parameters}
\label{sec3}

Effective temperatures ($T_{\rm eff}$) were estimated initially using the
Alonso, Arribas, \& Mart\'\i nez-Roger (1996a) calibrations versus $V-K$ and
$b-y$ colors, which were derived applying the Infrared Flux Method (Blackwell
et al. 1990), and cover a wide range in spectral types and metal content. Both
stars, BD $+23$\arcdeg 3130 and HD 84937, belong to a large sample of
metal-poor stars for which Garc\'\i a L\'opez et al. (1998) have derived
beryllium abundances. All the details on photometric measurements, as well as
on the selection of metallicity values for these two stars are provided in that
work. 

Surface gravities ($\log g$) were determined by comparing the observed
Str\"omgren $b-y$ and $c_1$ indices with synthetic ones generated using the
corresponding filter transmissions and a grid of Kurucz (1992) blanketed model
atmospheres fluxes. This is also explained in detail by Garc\'\i a L\'opez et
al. (1998), where a comparison is given with the spectroscopic estimates by
other authors. In particular, Nissen, H$\o$g, \& Schuster (1997) have recently
derived surface gravities for a sample of 54 metal-poor stars using {\it
Hipparcos} parallaxes to determine luminosities. For ten stars in common, the
gravities derived by Garc\'\i a L\'opez et al. are slightly lower
systematically, with a mean difference of $0.22\pm 0.13$ dex in $\log g$. As we
will see below, a  difference of this magnitude does not affect the boron
abundances derived from  the \ion{B}{1} $\lambda$ 2497 \AA\ lines in metal-poor
stars. BD $+23$\arcdeg 3130 had been generally considered in the past  as a
subgiant star with $\log g\sim 3-3.5$, but the comparison of its Str\"omgren
indices with our synthetic $uvby$ photometry indicates that it is a giant star
with $\log g\sim 2.5$. 

Garc\'\i a L\'opez et al. (1998) computed synthetic spectra in the region
around the \ion{Be}{2} doublet at $\lambda$ 3131 \AA\ (used to derive beryllium
abundances) for different combinations of the stellar parameters, within the
error bars provided by the photometric calibrations, and Table 1 lists the
values which best reproduce the observations. These values also reproduce 
well the \ion{B}{1} region observed in BD $+23$\arcdeg 3130 and HD 84937, with
the exception of the derived metallicity which is slightly higher than
that found from  optical and near-UV analyses, as explained in the next
section. 

Apart from our observations, we have also derived boron abundances for
Procyon and for the other metal-poor stars with available measurements of the
\ion{B}{1} $\lambda$ 2497 \AA\ doublet. For Procyon we choose $T_{\rm
eff}=6750$ K, $\log g=4.0$ and [Fe/H]$=0.0$, to be consistent with the
analysis of Duncan et al. (1997b; see below). For consistency, we also estimated
the stellar parameters for the other stars by the methods applied to our
stars. Most of them were studied by Garc\'\i a L\'opez et al. (1998). All
metal-poor stars  considered for boron abundances analysis in this work are
listed in Table 1. 


\section{Abundances Analysis}
\label{sec4}

\subsection{Oscillator Strengths}
\label{sec4.1} 

Due to the general lack of accurate oscillator strengths for the lines in the
\ion{B}{1} region, we used a line list computed by comparing LTE synthetic
spectroscopy with observed spectra of some stars. The initial line list was
taken from Duncan et al. (1997b) who analyzed dwarfs and giants of
approximately solar metallicity. The list was drawn from the Kurucz (1993)
LOWLINES list with wavelengths measured in the laboratory but the $\log gf$
values were adjusted to reproduce the spectra of Procyon, the Sun, $\alpha$ Cen
A \& B, and the metal-poor star HD 140283.  Several \ion{Fe}{1} lines predicted
by Kurucz as well as other lines of the same element studied by Johansson \&
Cowley (1988) and Johansson (1992) were added.  Duncan et al.'s (1997b) line
list is very similar to that compiled by Duncan et al. (1992).

Synthetic spectra were obtained by us using the code WITA2, a UNIX-based
version of ABEL6 code (Pavlenko 1991), which computes LTE atomic and molecular
synthetic profiles for a given model atmosphere. We employed initially a grid
of models provided by Kurucz (1992), which are interpolated for given values of
$T_{\rm eff}$, $\log g$, and [Fe/H]. To test the behavior of the Duncan et al.
(1997b) line list using WITA2 and a high-resolution HST spectrum, we compared a
$\lambda /\Delta\lambda\sim 90000$ spectrum of Procyon (observed by Lemke,
Lambert, \& Edvardsson 1993) with synthetic spectra obtained using the stellar
parameters cited above, a  microturbulent velocity $\xi=1.0$ km s$^{-1}$, a
wavelength step of 0.009 \AA, and convolved with a gaussian with a FWHM of
0.027 \AA\ to  reproduce the observed spectral resolution. The  stronger
\ion{B}{1} line at $\lambda$ 2497.723 \AA\ is seriously blended with  other
lines, and the abundance determination relies only on the \ion{B}{1} line  at
$\lambda$ 2496.772 \AA, more isolated and weaker, which is blended with  lines
of \ion{Co}{1} and \ion{Fe}{1} at $\lambda\lambda$ 2496.708 and 2496.870  \AA,
respectively. O'Brian \& Lawler (1992) and Johansson et al. (1993) provide
accurate values (which differ by only 0.03 dex) for the oscillator strength of
this \ion{B}{1} line. To improve the fit of the synthetic to the observed
spectrum in the region surrounding the weaker \ion{B}{1} line, we varied the
$\log gf$s of 13 of the 20 lines present in the 2496.4--2497.4 \AA\ region
until a good fit was obtained. In particular, we adopted the $\log gf$ value
provided by O'Brian et al. (1991) for the \ion{Fe}{1} $\lambda$ 2496.533 \AA\
line (with an uncertainty of 9\%), the \ion{Fe}{1} line $\lambda$ 2496.933 \AA\
was not taken into account, and we included two weak \ion{Fe}{1} lines at
$\lambda\lambda$ 2497.153 and 2497.420 \AA\ from the list used by Duncan et al.
(1992), modifying also their $\log gf$s by 0.1 and 0.25 dex, respectively. 

Figure 2 shows the comparison between the observed spectrum of Procyon and five
synthetic spectra computed for different values of the boron abundance. The set
of $\log gf$s employed to compute them reproduces the spectrum of Procyon with
a similar quality to those illustrated by Lemke et al. (1993) and Duncan et al.
(1997b). The boron abundance which best matches the observed spectrum is a
value between log N(B)$=2.0$ and 2.1 (in the customary scale where log
N(B)$=$log (B/H)$+$12). The sensitivity of the observed feature to the boron
abundance of the computed line is  high, as can be seen in Figure 2$b$: changes
of 0.1 dex around log N(B)$\sim 2.0$ can be appreciated. Taking into account
the uncertainty in the continuum placement and a change of 50 K in the assigned
effective temperature, the boron abundance provided by this analysis is log
N(B)$=2.05\pm 0.20$. Since the possibility of unidentified lines located close
to the \ion{B}{1} $\lambda$ 2496.772  \AA\ line cannot be excluded, this
abundance should be considered strictly as an upper limit. 


Our LTE boron abundance for Procyon is 0.15 dex smaller and 0.15 dex higher
than those obtained by Duncan et al. (1997b) and Lemke et al. (1993),
respectively. In the latter case, the model photosphere employed was computed
using the MARCS code with the opacity sampling method (OSMARCS) as described by
Edvardsson et al. (1993). Using the LTE code SPECTRUM included in the Uppsala
Synthetic Spectrum Package, we  have computed synthetic spectra for Procyon
with the line list previously compiled and an OSMARCS model atmosphere. In this
case, the boron abundance which best matches the observed spectrum is between
log N(B)$=1.9$ and 2.0, just 0.1 dex smaller than the value obtained using a
Kurucz model. Figure 3 shows the fits to the observed spectrum using both sets
of model photospheres and a common boron abundance log N(B)$= 2.0$. It can
be seen that they are very similar, being the larger differences in the
wavelength range 2496.8 to 2497.2 \AA\ (very slight changes in the $gf$ values
used would be needed to obtain a similar good fit with the OSMARCS model), but
not affecting significantly the abundance derived from the \ion{B}{1} $\lambda$
2496.772 \AA\ line. Kurucz models used in our analyses were computed with a
mixing length parameter $\alpha =1.25$, and we have checked (by running the
program ATLAS9 using the UNIX-based version assembled by M. Lemke and
distributed through the CCP7 database\footnote{http://star.arm.ac.uk/ccp7/})
that these models do not include overshooting. OSMARCS models do not include
overshooting either and were computed with $\alpha =1.5$. A very good agreement
between spectra computed with both sets of models was found also by Garc\'\i a
L\'opez, Severino, \& Gomez (1995) for the analysis of the \ion{Be}{2}
$\lambda$ 3131 \AA\ region, showing negligible differences in beryllium
abundances (using MARCS  models treating the line blanketing by means of
opacity distribution functions and applying wavelength-dependent scaling
factors to the total H$^-$ opacity).


The very good agreement between these independent analyses in a metal-rich star
like Procyon gives us a certain level of confidence to proceed to derive boron
abundances in other stars. It should be noted, however, that gross similarities
among the analyses does not preclude the possibility of systematic errors.
Furthermore, the presence of greater uncertainties in the model photospheres 
for metal-poor stars (e.g. Gratton, Caretta, \& Castelli 1996) will constrain 
also the validity of abundances derived in those stars.

\subsection{BD +23\arcdeg 3130 and HD 84937}
\label{sec4.2}

\subsubsection{Spectral Synthesis}
\label{sec4.2.1}

Using our line list and the stellar parameters listed in Table 1, we computed
synthetic spectra of the two stars observed by us using both Kurucz and OSMARCS
models. Computations were made with a step of 0.024 \AA\ in wavelength and
convolved with a gaussian to account for the spectral resolution of the
observations. We had initially assigned a metallicity of $-2.9$ to BD
$+23$\arcdeg 3130 following the Be analysis by Garc\'\i a L\'opez et al.
(1998); however, a better overall fit to the \ion{B}{1} region is obtained when
values of [Fe/H]$=-2.6$ and $-2.7$ are used for Kurucz and OSMARCS models,
respectively. Figure 4 shows the comparison between observed and synthetic
spectra (computed for different boron abundances) of this star using a Kurucz
model. The major features surrounding the \ion{B}{1} $\lambda$ 2496.772 \AA\
line are reproduced, although they are not fully fitted. The sensitivity of the
synthetic spectra to boron abundance is seen in detail in Figure 4$b$. The boron
abundance of BD $+23$\arcdeg 3130 is log N(B)$=-0.3$, and changes of 0.1 dex
around this value are clearly appreciated. The spectrum computed with log
N(B)=$-3$ (no boron) shows much smaller features for both lines of the
\ion{B}{1} doublet, indicating that we have a positive detection of boron in
this star. The errors in the boron abundance associated with variations of $\pm
150$ K in $T_{\rm eff}$, $\pm 0.3$ dex in $\log g$, and $\pm 0.3$ dex in [Fe/H]
are $\pm 0.2$, 0, and $\pm 0.2$, respectively. The \ion{B}{1} line is too weak
to be sensitive to the uncertainty of the adopted microturbulence ($\xi =1.0$
km s$^{-1}$). Placement of the continuum was checked by computing a synthetic
spectrum in the 2499--2501 \AA\ region using the Duncan et al. (1997b) line
list at those wavelengths. At $\lambda$ 2499.96 \AA\ both synthetic and
observed spectra reach a value of 1. Taking into account the S/N and the
possible errors in normalizing the observed spectrum, an additional error of
$\pm 0.1$ dex can be considered. Combining all these uncertainties, the final
LTE boron abundance of BD $+23$\arcdeg 3130 is log N(B)$=-0.30\pm 0.30$.


The analysis carried out with the same stellar parameters (except [Fe/H]) using
OSMARCS models provided an abundance log N(B)$=-0.40$, showing the same
sensitivity to changes in boron abundance as well as similar uncertainties
related to the choice of the stellar parameters and continuum location. A
comparison between the best fits to the observed spectrum obtained with both
sets of models is shown in Figure 5.


A similar procedure was followed to obtain the boron abundance of HD 84937. A
comparison between observed and synthetic spectra computed with a Kurucz model
is shown in Figure 6. The \ion{B}{1} line is weaker for a given abundance in HD
84937 than in BD $+23$\arcdeg 3130 because boron is more ionized. The observed
spectrum is reproduced with a value between log N(B)$=0.2$ and $0.0$ (Figure
6$b$) but, at this S/N level and with the uncertainties present in the
parameters of the lines surrounding the \ion{B}{1} line, a fit to the synthetic
spectrum computed without boron (log N(B)$=-3$), or with other intermediate
boron abundance, cannot be ruled out. This means that log N(B)$=0.1$ must be
considered an upper limit to the boron abundance in this star. As in the
previous case, it was necessary to increase the adopted metallicity from
[Fe/H]$=-2.25$ to $-2.1$ to obtain an overall good fit. Figure 7 shows the
comparison between the fits to the observed spectrum obtained using Kurucz and 
OSMARCS models with the same stellar parameters and boron abundance. It can be
seen how the overall behavior is very similar in both cases, as well as the
upper limit at log N(B)$=0.1$. The uncertainties of $\pm 120$ K in $T_{\rm
eff}$, $\pm 0.3$ dex in $\log g$, $\pm 0.15$ in [Fe/H], and continuum placement
will change this value within $\pm 0.25$ dex. Taking these uncertainties into
account, Table 1 reflects a conservative upper limit log N(B)$<0.35$ for this
star.



Spectral lines appear to be broader in BD $+23$\arcdeg 3130 with respect to HD
84937. The instrumental profile should be the dominant contributor to the line
widths and this ought to be the same for the two stars. Additional broadening
would arise if BD $+23$\arcdeg 3130 were a spectroscopic binary such that a
weaker line was present slightly shifted to lower wavelengths. However, this
star does not appear as a binary or suspected binary star in the literature,
especially in the survey of proper motions of about 500 stars studied by Carney
et al. (1994). On the other hand, HD 84937 is listed as a suspected binary in
the catalog of Carney et al. and classified as a certain binary by Stryker et
al. (1985) from radial velocity measurements, although is not considered as a
possible binary by Nissen et al. (1997). Duncan et al. (1997b) observed two
giant K-type stars belonging to the Hyades, as well as the field giant $\beta$
Gem. The composite spectrum of the Hyades stars and the spectrum of $\beta$ Gem
show similar broad lines which are, in general, deeper than those observed in
BD $+23$\arcdeg 3130 due to the big differences in metallicity. In any case,
the spectral syntheses carried out by us and by Duncan et al. (1997b) show a
very high sensitivity of the \ion{B}{1} feature to the boron abundance in giant
stars.

\subsubsection{Stellar Fluxes}
\label{4.2.2}

A complementary test to check the validity of the model photospheres employed
in the analysis consists in comparing predicted and observed stellar fluxes at
the wavelength region under study. This is especially important in the
ultraviolet, where an enormous number of lines are present such that the 
models may not account for the line blanketing. We exploit the absolute flux
calibration provided with the HST spectra by following the Duncan et al.'s
(1992) procedure, using the stellar magnitude at the $K$ (2.2 $\mu$m) band to
normalize the fluxes. $K$ magnitudes for our stars (6.98 for BD $+$23\arcdeg
3130 and 7.10 for HD 84937) were taken from Laird, Carney, \& Latham (1988).
Theoretical fluxes corresponding to the Kurucz models were obtained by
interpolating the grid of models currently available at the CCP7 database,
while they were computed individually for the OSMARCS models used in the
analyses.

Below $\lambda$ 4500 \AA\ OSMARCS models are computed using opacity sampling
techniques, while at longer wavelengths opacity distribution functions (ODF)
are employed. The latter method is used for computing Kurucz models at all
wavelengths. Comparisons between both set of models show a very good agreement
for IR wavelengths ($\lambda\gtrsim 1$ $\mu$m), which decreases slightly with
decreasing wavelength, and a significant difference appears for $\lambda <
4500$ \AA. Figure 8 shows the comparison in the UV (surrounding the \ion{B}{1}
region) for the Kurucz and OSMARCS models used for computing the synthetic
spectra of Figure 7. It can be seen in the upper panels the differences
associated with both ways of considering the line blanketing. These differences
are strongly reduced by degrading the OSMARCS fluxes to the same resolution as
Kurucz ones. That is shown in the lower panels, where we have convolved the
OSMARCS fluxes with a gaussian with a sigma of 7 \AA. This behavior is
consistent with the very small differences observed between their synthetic
spectra, and indicates that the continuum fluxes are independent of model
family.


Comparisons between observed and predicted fluxes were carried out at $\lambda$
2500 \AA. The ratios of the observed to predicted fluxes obtained using OSMARCS
models for BD $+23$ 3130 and HD 84937 are 1.24 and 1.20, respectively. These
values are larger than the mean ratio $0.98\pm 0.12$ resulting from the three
metal-poor stars analyzed by Duncan et al. (1992). However, a ratio of 1.2 is
compatible with the combined uncertainties in the models and in the HST flux
calibrations, and we conclude that the models correctly reproduce the flux at
$\lambda$ 2500 \AA.

\subsection{Other Metal-Poor Stars}
\label{sec4.3}

With the aim of tracing the evolution of boron from the early times of the
Galaxy to the present, we decided to include in our study all metal-poor
stars observed for their boron abundances. These stars are analyzed in a
recent work by Duncan et al. (1997a), including their own observations as
well as the three stars observed  by Duncan et al. (1992).

\subsubsection{Spectral Synthesis}
\label{sec4.3.1}

Rebull et al. (1996) derived the boron abundance in a very interesting
metal-poor star: BD $-13$\arcdeg 3442.  These authors found an LTE abundance of
log N(B)$=-0.06\pm 0.25$, considering the following stellar parameters:
[Fe/H]$=-2.96$, $\log g =3.0$, and $T_{\rm eff}=6100$ K. This value was revised
to $0.01\pm 0.20$ by  Duncan et al. (1997a, 1998)  using slightly different
parameters ([Fe/H]$=-3.00$, $\log g =3.75$, and $T_{\rm eff}=6250$ K). Given
that we had found only an upper limit to the abundance of the less metal-poor
star HD 84937 with similar $T_{\rm eff}$, we decided to carry out the analysis
of BD $-13$\arcdeg 3442 using our tools. We took the observed spectrum from
Figure 2 of Rebull et al. (1996) and computed synthetic spectra with different
boron abundances using the same stellar parameters. The fit obtained is very
similar to that of Rebull et al. (1996) and the boron abundance derived is log
N(B)$=0.0$. With respect to the sensitivity of the synthetic spectra to the
abundance, we found the same behavior as in HD 84937, i.e., there are very
small differences between spectra computed within a range of three orders of
magnitude in boron abundance. In order to have the abundance for this star on 
our $T_{\rm eff}$ scale, we computed new spectra using  the stellar parameters
listed in Table 1, but increasing the metallicity by 0.1 dex to obtain a better
overall fit (with both Kurucz and OSMARCS model atmospheres). Our best estimate
for the boron abundance of this star is an upper limit at $-0.1$ dex, and
considering the  usual uncertainties in stellar parameters, S/N, and continuum
placement this  value could change by $\pm 0.30$ dex. The adopted upper limit
taking into account this uncertainty is then log N(B)$<0.20$.

A similar behavior can be seen for the star BD $+26$\arcdeg 3578 in Figure 2$d$
of Duncan et al. (1997a). There is but a very small difference between their
synthetic spectra computed with log N(B)=$-0.13$ and without boron.  Our
analysis of this star gives an upper limit of log N(B)$<0.00$, using our
adopted $T_{\rm eff}$ and $\log g$ but increasing the metallicity to
[Fe/H]$=-2.25$ and $-2.30$ for Kurucz and OSMARCS models, respectively (instead
of the value $-2.5$ listed in Table 1). The upper limit adopted taking into
account the uncertainties in the analysis is log N(B)$<0.25$. 

These spectral syntheses show that the combination of high effective
temperature and low abundance makes it difficult to obtain accurate estimates
of boron in metal-poor stars with $T_{\rm eff}\gtrsim 6000$ K using the
\ion{B}{1} $\lambda$ 2496.772 \AA\ line. Therefore, we  consider  the available
spectra of these stars (including also BD $+3$\arcdeg 740, for which we have
not computed synthetic spectra) to provide upper limits to the  boron
abundances. This consideration removes four important objects (in terms of
their low metallicities) from the sample used to delineate the boron
evolution. 

A systematic effect present in all our spectral syntheses is that the
metallicity indicated by the far UV (boron) spectra seems to be higher than
that needed to reproduce the near UV (beryllium) and optical regions. In two
cases (HD 84937 and BD $-13$\arcdeg 3442) these differences are within the
error bars of the adopted metallicities, in another case (BD $+23$\arcdeg 3130)
the difference is 0.1 dex higher than the error bar, and for BD $+26$\arcdeg
3578 this difference is 0.10--0.15 dex larger than the adopted error bar. A
poor fit to the metallic lines surrounding the \ion{B}{1} region is also
observed in the spectral synthesis provided by Rebull et al. (1996) for BD
$-13$\arcdeg 3442. 

Spectral lines in the \ion{B}{1} region are not resolved with the resolution of
these GHRS observations. A line like the \ion{Fe}{1} $\lambda$ 2496.533 \AA\ is
very deep and saturated, and thus the core of the line is formed at the top of
the atmosphere. We have checked that the model atmospheres are sufficiently
extensive that the line is computed correctly. There remains the question as to
whether the model represents adequately the real upper photosphere. Errors in
the UV opacity, departures from LTE, or effects of convective overshoot and
thermal inhomogeneities might contribute to the fact that the far UV synthetic
spectra seem to need a higher metallicity to reproduce the observations.
Although there is no clear correlation between these differences and the
adopted metallicities for the four stars considered, this effect could also be
metallicity dependent and so could cause a slight change of slope in the
boron--[Fe/H] relation discussed in this paper.

\subsubsection{Equivalent Widths}
\label{sec4.3.2}

To adapt the abundances given by Duncan et al. (1997a) to our stellar
parameters for those stars for which we have not carried out a detailed
spectral synthesis, we computed the \ion{B}{1} $\lambda$ 2496.772 \AA\ line
with the boron abundances and stellar parameters listed in that work using
Kurucz model atmospheres, and measured the corresponding synthetic equivalent
widths. The adopted LTE abundances were those which matched these equivalent
widths with the stellar parameters listed in Table 1. The sensitivity of our
spectral synthesis to the stellar parameters is similar to that carried out by
Duncan et al. (1997a), and we adopted their final errors to the abundances.
Table 1 shows the final abundances obtained in this way. The differences
between LTE boron abundances obtained using our stellar parameters and those of
Duncan et al.  (1997a) are in the range $-0.12$ to 0.08 dex for all the stars,
except for BD $+3$\arcdeg 740 with a difference of 0.25 dex. These differences
are mainly due to differences in $T_{\rm eff}$ (up to 170 K for BD $+3$\arcdeg
740), and for stars with low metallicity the changes in abundance are larger
for a given difference in $T_{\rm eff}$ with respect to more metal-rich stars.

We also derived LTE abundances for these stars using OSMARCS models and the
equivalent widths previously computed with Kurucz model atmospheres. These
abundances are also listed in Table 1. As expected, there is a systematic
slight difference between both sets of abundances with a mean value $0.05\pm
0.03$. In all abundance estimates we have used a microturbulent velocity of 1
km s$^{-1}$. For the most metal-poor stars analyzed the synthetic spectra have
shown that the dependence of boron abundance on the microturbulence is
negligible. In the most metal-rich stars of the sample, with higher boron
abundances, the effect of changing the microturbulence by one additional km 
s$^{-1}$ induces a change of 1--5\% in the equivalent widths of the
\ion{B}{1} lines. Changes of this order affect the boron abundance by
0.01--0.07 dex for stars with $-2.2<$[Fe/H]$<-0.5$. 

As explained in Section 3, effective temperatures were estimated using the
$T_{\rm eff}-(V-K)$ and $T_{\rm eff}-(b-y)$ calibrations of Alonso et al.
(1996a), and corrected afterwards until a good fit to the \ion{Be}{2} $\lambda$
3131 \AA\ region was obtained (Garc\'\i a L\'opez et al. 1998). With this
procedure, two of the stars considered in this paper (BD $+3$\arcdeg 740 and HD
140283) show $T_{\rm eff}$ values with significant differences with respect to
those obtained using other calibrations. For BD $+3$\arcdeg 740 the temperature
is directly a mean value between 6180 and 6410 K from $V-K$ and $b-y$ colors,
respectively. However, given that it is hotter than 6000 K we are considering
its boron measurement as an upper limit and it is not relevant for the
discussion in the next Section. The temperature adopted for HD 140283 is lower
than other values usually quoted in the literature. Using e.g. $T_{\rm
eff}=5690$ K (Carney 1983; Alonso, Arribas, \& Mart\'\i nez-Roger 1996b), the
LTE boron abundance for this star would rise by about 0.05 dex only.

\subsection{NLTE Boron Abundances}
\label{sec4.4}

Kiselman (1994) showed that the approximation of line formation in local
thermodynamic equilibrium leads to significant errors in abundance
determinations for solar-type metal-poor stars using \ion{B}{1} lines. These
lines are especially susceptible to NLTE effects because the overexcitation and
overionization effects work in the same direction in making the lines weaker
than if Boltzmann and Saha equilibria were valid. This is not the case for the
\ion{Be}{2} lines that are used for beryllium abundance analysis -- here
overionization and overexcitation work in opposite directions on the line
strengths as was shown by Garc\'\i a L\'opez et al. (1995) and Kiselman \&
Carlsson (1995). The NLTE behavior of lithium is more complicated, partly
because the generally higher lithium abundances make the lithium atoms
susceptible for various NLTE effects typical for strong lines. Carlsson et al.
(1994) have published NLTE abundance corrections for an extensive grid of
photospheric parameters, and Pavlenko et al. (1995) and Pavlenko \& Magazz\`u
(1996) provide NLTE curves of growth which extend to very low effective
temperatures.

Kiselman \& Carlsson (1996) provide NLTE corrections for LTE boron abundances 
computed using OSMARCS models. These should in principle not be applied to LTE
abundances acquired with other models. The fact an LTE analysis of a line
gives similar results for two different models is not in itself a guarantee
that the NLTE analysis would. Line formation in NLTE involves coupling 
between different frequency regions and thus a more complicated dependence
on the photospheric temperature and pressure structures and the background 
opacities than what is the case in LTE. The different treatment of UV opacities
(see our discussion on stellar fluxes) makes it possible that consistent
NLTE line modeling using the Kurucz models could give results differing from
the OSMARCS results, even though the LTE abundances are so similar.

In order to be fully consistent, we choose as our final estimates of boron
abundances our OSMARCS LTE results with the NLTE corrections of Kiselman \& 
Carlsson applied. Correcting  the LTE abundances obtained with Kurucz models
using the Kiselman \& Carlsson prescriptions would mean that the final
abundances would be formally on the OSMARCS NLTE scale and slightly
overestimated.  The resulting abundances are given in Table 1. We will adopt
for the NLTE abundances the same error bars derived in the LTE analysis. The
sensitivity on uncertainties in stellar parameters is not significantly
different in LTE and NLTE. The dependence on atomic data is much more difficult
to asses (especially on $\log gf$), and the sensitivity on other data is of
second order. Furthermore, possible systematic errors are not easy to estimate.
The errors in the stellar radiation fields are probably the most pertinent ones
which bother the NLTE corrections, but we are considering that the continuous
opacities employed in the OSMARCS models reproduce the observed fluxes. 

\section{Discussion}
\label{sec5}

\subsection{HD\,84937 and BD +23\arcdeg 3130}
\label{sec5.1}

HD 84937 was selected as a key target for this work because it is the only star
for which there exist previous measurements of $^{6,7}$Li and $^9$Be. However,
due to its high $T_{\rm eff}$, our observations failed to provide a convincing
detection of the \ion{B}{1} $\lambda$ 2496.772 \AA\ line and consequently we
were forced to set an upper limit to the boron abundance. This disappointment
led us to reexamine published spectra of similarly warm extremely metal-poor
stars and to conclude that upper limits to rather than certain determinations 
of the boron abundance may be reached. This result has some influence on
determinations of the evolution of  the boron abundance with metallicity.

In the case of BD $+23$\arcdeg 3130, a cool giant, a concern may be that boron
has been depleted below its initial abundance. To investigate this concern, we
compare its Li and Be abundances with those of some other stars in our sample.
Equivalent widths of the \ion{Li}{1} $\lambda$ 6708 \AA\ line were taken from
Thorburn (1994) and Ryan et al. (1996), and the abundances were derived using
the stellar parameters adopted in Table 1 and NLTE curves of growth provided by
Ya. V. Pavlenko (obtained as described in Pavlenko et al. 1995 and Pavlenko \&
Magazz\`u 1996). Beryllium abundances were derived by Garc\'\i a L\'opez et al.
(1998). Li and Be abundances are listed in Table 2, where the latter include an
upwards correction of 0.1 dex to consider the mean difference with the 
gravities derived by Nissen et al. (1997) from {\it Hipparcos} parallaxes
(which does not affect either boron or lithium abundances). 


Evidently but not surprisingly Li is depleted in BD $+23$\arcdeg 3130  by more
than 0.7 dex. A part of the observed depletion may have occurred as a dilution
when the star evolved to the giant phase. Boron and beryllium are less fragile
elements than lithium, and it is unlikely that  their abundances have been 
reduced by nuclear reactions. It cannot be completely ruled out that these
abundances have been slightly reduced by dilution. It will be shown below that
the Be and B abundances of the star fit the trends with metallicity defined by
the other stars, and a small increment of the observed abundances of BD
$+23$\arcdeg 3130  would not change the general conclusions about the synthesis
of boron.

\subsection{Evolution of Boron with Metallicity}
\label{sec5.2}

Initial results on the evolution of the B abundance with metallicity ([Fe/H])
were provided by Duncan et al. (1992) from just three LTE B abundances of halo
stars. The linear trend between log N(B) and [Fe/H] was confirmed in essence 
by Kiselman \& Carlsson (1996) who included non-LTE effects for boron, and by
Duncan et al. (1997a) who analyzed additional stars. Our rediscussion adds two
stars and  recognizes that only upper limits to the B abundance are presently
obtainable for the warmest  stars. In addition, we give B and Be abundances for
stellar parameters derived in a uniform way for the entire sample.

Figure 9 shows the LTE and NLTE boron abundances of those stars listed in Table
1 against [Fe/H]. In the upper panel the symbols are also labeled with the
adopted effective temperatures. As explained in Section 4, we are only
considering as real measurements those made in metal-poor stars with $T_{\rm
eff}< 6000$ K. NLTE corrections also affect more strongly to the hotter stars,
inducing a larger scatter in the boron-metallicity plot  when including the
abundances associated with the upper limits. For comparison, we have also
included in Figure 9 the boron abundances of two solar-metallicity stars:
$\iota$ Peg and $\theta$ UMa, taken directly from Lemke et al. (1993). NLTE
effects on the B abundances were  taken into account for these stars following
Kiselman \& Carlsson (1996).


Two sets of fits are shown in Figure 9. Dashed lines represent linear
regressions performed using all metal-poor stars with $T_{\rm eff}< 6000$  K,
and solid lines restrict the fits to cool stars with [Fe/H]$< -1$. This
distinction is done to  compare with the behavior of beryllium abundances, for
which a change of slope at [Fe/H]$\sim -1$ is apparent from the observations
(Garc\'\i a L\'opez et al. 1998). All regression analyses were carried out
taking into account the error bars in boron abundances and metallicities, using
the routine {\it fitexy} of Press et al. (1992) available  in the IDL Astronomy
User's Library. These are the results:

\begin{itemize}

\item LTE, $T_{\rm eff}< 6000$ K:

$\log{\rm N(B)}=2.63(\pm 0.18)+1.03(\pm 0.10)\times {\rm [Fe/H]}, 
\, n=9, \, \chi^2_\nu=0.86, \, q=0.54$,

where $n$ is the number of stars included in the fit, $\chi^2_\nu$ is $\chi^2$
divided by the number of degrees of freedom $\nu$ (also called {\it reduced}
$\chi^2$;  $\nu =n-2$ in this case), and $q$ is the probability that a correct
model would give a value equal or larger than the observed  chi squared (a
small value of $q$ indicates a poor fit; $q>0.1$ provides an acceptable fit).

\item LTE, $T_{\rm eff}< 6000$ K and [Fe/H]$< -1$:

$\log{\rm N(B)}=3.06(\pm 0.31)+1.23(\pm 0.16)\times {\rm [Fe/H]}, 
\, n=7, \, \chi^2_\nu=0.42, \, q=0.84$.

\item NLTE, $T_{\rm eff}< 6000$ K:

$\log{\rm N(B)}=2.65(\pm 0.17)+0.89(\pm 0.10)\times {\rm [Fe/H]}, 
\, n=9, \, \chi^2_\nu=0.86, \, q=0.54$.

\item NLTE, $T_{\rm eff}< 6000$ K and [Fe/H]$< -1$:

$\log{\rm N(B)}=3.07(\pm 0.29)+1.08(\pm 0.15)\times {\rm [Fe/H]}, 
\, n=7, \, \chi^2_\nu=0.34, \, q=0.89$.

\end{itemize}

\noindent The plots make evident that a linear increase of the boron abundance
with increasing metallicity describes very well the trend observed in the data,
and  a change of slope may occur at the halo-disk transition at [Fe/H] $\sim
-1$.  

Duncan et al. (1997a) found slopes of $0.96\pm 0.07$ and $0.70\pm 0.07$ for
their linear fits to 11 stars in LTE and NLTE, respectively. Our corresponding 
slopes of $1.02\pm 0.10$ and $0.87\pm 0.10$, obtained using 9 stars (without
upper limits) are consistent with those of Duncan et al. within the  error
bars. If we include the data associated with the metal-rich stars $\iota$ Peg
and $\theta$ UMa, which have not been adjusted to our $T_{\rm eff}$ scale, the
LTE and NLTE slopes do not change appreciably ($0.96\pm 0.08$ and $0.85\pm
0.08$, respectively). The slopes are slightly steeper and outside the error
bars when compared with the results of Duncan et al., if we consider only those
7 stars with [Fe/H] $< -1$. We will consider our NLTE abundances and the fit
for stars with $T_{\rm eff}< 6000$ K and [Fe/H]$< -1$ as the actual values to
be used in the discussion performed below this point.

In studies of boron evolution, oxygen abundances may be a preferable
alternative to the iron abundances for astrophysical reasons. In particular,
oxygen is synthesized primarily by the most massive supernovae whose ejecta
(see below) may be the source of the energetic particles responsible for
boron production by spallation. Unfortunately, observational determinations of
oxygen abundances in metal-poor stars have been a matter of controversy during
the last years (e.g. Abia \& Rebolo 1989; Bessell, Sutherland, \& Ruan 1991;
Nissen et al. 1994; Cavallo, Pilachowski, \& Rebolo 1997). We examined the
boron-oxygen trend using the oxygen abundances found by Cavallo et al. (1997)
from high-resolution spectra of the \ion{O}{1} IR triplet in a sample of 24
stars with $-2.96\leq [{\rm Fe/H}]\leq -0.48$. These results were employed to
define the relation between [O/H] and [Fe/H] which was then used to establish
the B versus O trend. We find:

$\log{\rm N(B)_{NLTE}}=2.71+1.47\times {\rm [O/H]}$.

\noindent This slope is steeper than the comparable trend found by Duncan et
al. (1997a), with a slope of 0.82$\pm 0.10$ when using a mean [O/H] versus
[Fe/H] relation established using abundances extracted from the literature for
the stars in their sample. Almost in parallel with our analysis of boron,
Israelian et al. (1998) are deriving oxygen abundances in a wide sample of
metal-poor stars, including those considered here, using OH lines in the
3085--3256 \AA\ region and the same stellar parameters employed in this paper.
Until controversy about oxygen in metal-poor stars is clearly resolved, it must
suffice to note that the boron and oxygen abundances are approximately linearly
related.

\subsection{Boron and Beryllium}
\label{sec5.3}
 
Two modes of nucleosynthesis of boron compete (in theory) for primacy:
spallation reactions driven by energetic (cosmic rays) particles in
extra-stellar environments, and neutrino-induced spallation of $^{12}$C
occurring in Type II supernovae.  A measure of the relative contributions of
the two modes is offered by the B/Be ratio (at a given [Fe/H]) because
beryllium is synthesized (in theory) only by spallation reactions with high
energy particles. Moreover, in these spallation reactions, the B/Be production
ratio is essentially independent of the composition and the energy spectrum of
the high energy particles. Thus, the observed B/Be ratio reveals the relative
roles of the two modes of boron nucleosynthesis.

Figure 10 shows the comparison between boron (NLTE) and beryllium abundances
against metallicity. Beryllium data were taken from Garc\'\i a L\'opez et al.
(1998), who carried out a detailed spectral synthesis of the \ion{Be}{2}
$\lambda$ 3131 \AA\ region for 28 stars with $-0.4\leq [{\rm Fe/H}]\leq -3$,
using high-resolution observed spectra. These authors also included all
metal-poor stars with accurate Be abundance available, and  repeated
measurements on the same star were averaged both in Be and in [Fe/H]. Also
shown for comparison are the Be abundances given by Lemke et al. (1993) for the
metal-rich stars $\iota$ Peg and $\theta$ UMa, estimated from Boesgaard's
(1976) equivalent widths measured on photographic spectra. A linear regression
analysis for stars with [Fe/H] $< -1$, without considering the upper limits
(not shown in Fig. 10), provides:

$\log{\rm N(Be)}=1.96(\pm 0.17)+1.15(\pm 0.09)\times {\rm [Fe/H]}, 
\, n=29, \, \chi^2_\nu=0.81, \, q=0.75$.

\noindent These values of $\chi^2_\nu$ and $q$ indicate an excellent fit over
a significant sample of metal-poor stars. These are LTE abundances but 
corrections for non-LTE effects are small ($\leq 0.1$ dex in general --see
Garc\'\i a L\'opez et al. 1995 and Kiselman \& Carlsson 1995--). 


The slopes of the observed trends against metallicity are very similar for Be
and B for stars with [Fe/H] $< -1$. Indeed, the slopes are identical within the
error bars.  Duncan et al. (1997a) found a similar behavior when comparing
their NLTE B abundances and published Be abundances for their sample of stars.

Using the Be abundances given by Garc\'\i a L\'opez et al. (1998),  derived
with the stellar parameters listed in Table 1 and shown in Table 2, the
average  B/Be ratio for  the seven stars with $T_{\rm eff}< 6000$ K and [Fe/H]
$< -1$ is B$_{\rm NLTE}$/Be $=20\pm 10$ (1 $\sigma$). For all of these stars,
with the  exception of BD $+23$\arcdeg 3130, independent beryllium measurements
are available in the literature based on high-resolution CCD spectra,  slightly
different stellar parameters, and employing  spectral synthesis or curves of
growth. Garc\'\i a L\'opez et al. (1995) showed that these  independent LTE
analyses provide very similar Be abundances when adjusted for differences in
the stellar parameters (especially the surface gravity). The average Be
abundances for these seven stars differ by $0.10\pm 0.04$ dex  from  the
individual values listed in Table 2. Using the average Be values we obtain a
mean ratio B$_{\rm NLTE}$/Be $=17\pm 10$. Figure 11 shows the  behavior of log
(B$_{\rm NLTE}$/Be) against metallicity. The solid line represents the mean
value and dotted lines show the $1 \sigma$ uncertainties:  $\log ({\rm
B_{NLTE}/Be})=1.31\pm 0.21$. All ratios are compatible with this mean value and
its uncertainty within the error bars. A linear regression fit taking into
account the errors in both coordinates provides:

$\log ({\rm B_{NLTE}/Be})=1.27(\pm 0.45)-0.01(\pm 0.22)\times {\rm [Fe/H]}, 
\, n=7, \, \chi^2_\nu=0.32, \, q=0.90$,

\noindent which is fully consistent with the mean value and is shown as a
dashed line in Figure 11. Duncan et al. (1997a) obtained a mean value B$_{\rm
NLTE}$/Be $=15$ using 9  stars. Be abundances adopted by Duncan et al. (1997a)
are, in general, larger than those adopted here and that is the main reason why
they obtain a smaller B/Be ratio. 


It is worthwhile to consider the trends given by the regression analyses of
boron and beryllium against metallicity, which better represent their overall
behavior. Molaro et al. (1997) derived beryllium abundances for a sample of
metal-poor stars  and, combining them with measurements available in the
literature, obtained a linear fit of Be abundance against metallicity with a
slope of $1.07\pm 0.08$ for 19 stars in the interval $-2.7 < [{\rm Fe/H}]<
-0.8$. This value is compatible with the slope $1.15\pm 0.09$ found by Garc\'\i
a L\'opez et al. (1998) for 29 stars in the interval $-3 < [{\rm Fe/H}]< -1$.
Using this relation and the fit obtained in this work between log N(B)$_{\rm
NLTE}$ and [Fe/H] for stars with $T_{\rm eff}< 6000$ K and [Fe/H] $< -1$, we
find a mean ratio of B$_{\rm NLTE}$/Be $=18$. Using our boron relation and the
beryllium fit provided by Molaro et al. (1997), this ratio is just 17.

\subsection{Nucleosynthesis}
\label{sec5.4}

Two observational facts about the B and Be abundances stand out: (i) metal-poor
([Fe/H]$\leq -1$) stars have a constant boron-to-beryllium ratio $\sim 20$, 
and (ii) the relation between boron (or beryllium) and metallicity has a slope
of $\sim 1$. The linear dependence, first seen for Be vs Fe by Gilmore et al.
(1992) and then by Duncan et al. (1992) for B vs Fe, was a surprise. Simple
models of spallation of ambient CNO nuclei in the interstellar medium by cosmic
ray protons and $\alpha$s had predicted a quadratic dependence (B $\propto$
Fe$^2$). Several explanations for the observed linear dependence have been
advanced. In essence, most explanations invoke spallation of CNO nuclei in
energetic collisions with ambient protons and $\alpha$s. If the composition of
the energetic particles is greatly enriched in CNO, as could be assumed for
supernovae ejecta or winds from Wolf-Rayet stars, a quasi-linear relation
between B (and Be) and metallicity would result (Duncan et al. 1992; Vangioni-
Flam et al. 1996). Other possibilities, such as local spallation by high-energy
protons and $\alpha$s in enriched regions around supernovae, seem less probable
(Feltzing \& Gustafsson 1994). At any rate, fact (ii) is to be explained
largely by astrophysical issues.

The simplest interpretation of fact (i) is that B and Be are made by the same
nucleosynthetic process. This process is spallation of CNO nuclei by collisions
with protons and $\alpha$s. Early calculations of yields from spallation gave a
B/Be ratio near 20: Reeves (1994) gives the ratio as 17 for production by high
energy ($\geq$ 100 MeV nucleon$^{-1}$) cosmic rays. The primary factors setting
the ratio are the cross-sections for the spallation reactions, i.e., nuclear
physics rather than astrophysics. Theoretical yields are thoroughly discussed
by Ramaty et al. (1996) who investigate their dependence on the energy spectrum
and composition of the energetic particles (cosmic rays), and on the
composition of the ambient medium.  This investigation shows that a ratio of
B/Be $\sim 20$ from spallation does not constrain the composition of either the
energetic particles or the ambient medium, and, provided the  energy spectrum
of the particles is not biased to threshold energies for the proton and
$\alpha$ on CNO nuclei spallation reactions, the ratio does not constrain the
energy spectrum. Ramaty et al. stress that the various combinations of energy
and composition of energetic particles and composition of the ambient medium
call for different amounts of energy to be put into energetic particles:
certain combinations are very inefficient. If the particles have
the predicted composition of the (most massive) supernovae ejecta or Wolf-Rayet
winds, Ramaty et al.'s condition on the energy available for high energy
particles is met. The condition is very difficult to impossible to meet if the
high energy particles have the composition of the Sun, the galactic cosmic rays
or of the ejecta of low mass Type II supernovae. It is intriguing that this
argument about the composition of the high energy particles is consistent with
that adduced from the linear relation between B (and Be) and Fe. This constancy
of Be/Fe and B/Fe for metal-poor stars led Ramaty et al. (1997) to relate the
production rates of Be and B to the Fe production rate by Type II supernovae
via cosmic-ray acceleration. Considering that the $^{11}$B/$^{10}$B value in
metal-poor stars is probably not lower than the solar system ratio of $4.05\pm
0.2$ (Chaussidon \& Robert 1995), a B/Be $\sim 20$ would imply an initial
acceleration energy of about 60 MeV nucleon$^{-1}$ (see Fig. 13$a$ of Ramaty et
al. 1997), and a rapid increase in the initial energy needed ($>100$ MeV
nucleon$^{-1}$) for slightly lower values (17--18) of B/Be.

Beryllium is thought to be synthesized only by spallation reactions but boron
synthesis by spallation may be supplemented by Type II supernovae in which
boron is made by neutrino-induced spallation of carbon (the $\nu$-process). 
Timmes, Woosley, \& Weaver (1996) in a detailed study of galactic chemical
evolution suggest that predicted yields of boron by supernovae reproduce well
the B versus Fe trend of the halo and disk {\it without} the introduction of
spallation processes. This is incompatible with the attribution of Be synthesis
to spallation processes and the observed ratio B/Be $\sim 20$ which implies
that boron synthesis is also by spallation processes.  We conclude that yields
of boron from the $\nu$-process have been overestimated. Vangioni-Flam et al.
(1996), on the basis of few published B abundances, concluded that  the
$\nu$-process is at least a factor of 5 less efficient than supposed by Timmes
et al. (1995).

Alternative models of chemical evolution of the Galactic halo (e.g., Casuso \&
Beckman 1997) also give consistent results with the observed linear trends of B
and Be vs. Fe at very low metallicity. These authors consider an exponentially
increasing flow of gas from the halo to form the bulge. The strong Be and B
enrichment in the halo is attributed to the presence of more star-forming gas
at very low metallicity, and to increased yields of carbon in intermediate and
low-mass metal-poor stars. The Be/B ratios that they predict are also similar
to the observed ones. It seems that under certain hypothesis for the evolution
of the gas and for the Initial Mass Function, conventional cosmic ray
spallation processes may then provide an explanation to the observations.

In short, the ratio B/Be $\sim 20$ is strong evidence that B and Be are
products of spallation reactions between protons and/or $\alpha$s and CNO
nuclei. 

The beryllium abundances (Figure 10) indicate that there is a change of slope
in the run of Be with Fe at about [Fe/H] $\sim -1$, the canonical transition
between halo and disk (a more detailed discussion based on the Be data is
presented in Garc\'\i a L\'opez et al. 1998). The very limited data on the
boron abundance in disk stars suggests that the B versus Fe trend follows the
Be versus Fe trend such that B/Be remains constant at $\sim 20$. Additional
data for boron at these metallicities is crucial to confirm this and for a
better understanding of the halo-disk transition. The meteoritic abundances
(Grevesse, Noels, \& Sauval 1996) give B/Be $= 23 \pm 5$ in agreement with our
result from metal-poor stars.

\section{Conclusions}
\label{sec6}

BD $+23\arcdeg$ 3130 observed and analyzed here is the most metal-poor star for
which a firm measurement of the boron abundance presently exists (two slightly
more metal-poor stars -- BD $-13\arcdeg 3442$ and BD $+3\arcdeg$ 740 -- have
been analyzed but we consider that the \ion {B}{1} $\lambda$ 2497 \AA\ line has
not been positively identified in published spectra.) Our result and slightly
revised results from other metal-poor stars establish the run of boron with
iron for halo stars: the boron abundance increases linearly, with a slope $\sim
1$, with the iron abundance. With beryllium abundances consistently derived for
the same stars using the same atmosphere parameters, the B/Be ratio is found to
be $\sim 20$. 

A direct interpretation of these results is that B and Be in the halo were
synthesized by spallation reactions between ambient protons and $\alpha$s and
high energy particles with a composition enriched in CNO. The B/Be ratio is in
fine agreement with the predictions (Vangioni-Flam et al. 1996; Ramaty et al.
1996, 1997). This is a significant result because the  predicted B/Be ratio is
largely independent of the details of the modeling of the spallation process
being largely determined  by laboratory measurements of spallation
cross-sections. The linear trend of B (and Be) with Fe is dependent on the
astrophysics of spallation in the halo but can be reproduced with plausible
assumptions (Vangioni-Flam et al. 1996).

A consistent picture of boron synthesis has emerged for the halo of our Galaxy.
Stars more metal-poor than BD $+23\arcdeg$ 3130 are now known but it will not
be easy to extend measurements of boron abundance to these stars. The \ion
{B}{1} $\lambda$ 2497 \AA\ line will be too weak in the warmest of these stars,
as we have found for less extreme stars. Some extension to lower metallicity
will be possible for cool stars but for the fact that the candidate stars are
faint requiring prohibitively long exposures with HST. Almost certainly, the
boron plateau  signifying detection of primordial boron will elude us; boron
(and beryllium) from homogeneous  and inhomogeneous big bangs is predicted to
have an abundance 7 or 8 dex below the abundance found here for BD $+23\arcdeg$
3130 (Thomas et al. 1993, 1994).

\acknowledgements

It is a pleasure to acknowledge the following people for their help with the
analyses described in this work: Martin Asplund for computing several OSMARCS
models and their corresponding fluxes, Ya. V. Pavlenko for providing NLTE Li
curves of growth, and the staff at STScI and ECF for their support during all
phases of these HST observations. The comments of an anonymous referee have been
of value in improving the content of this article.

This work was partially supported by the Space Telescope Science Institute
(grant GO-05862), and the Spanish DGES under projects PB92-0434-C02-01 and
PB95-1132-C02-01. BE and BG acknowledge support from the Swedish Natural
Science Research Council (NFR) and the Swedish Space Board.

\newpage


\begin{figure}
\plotone{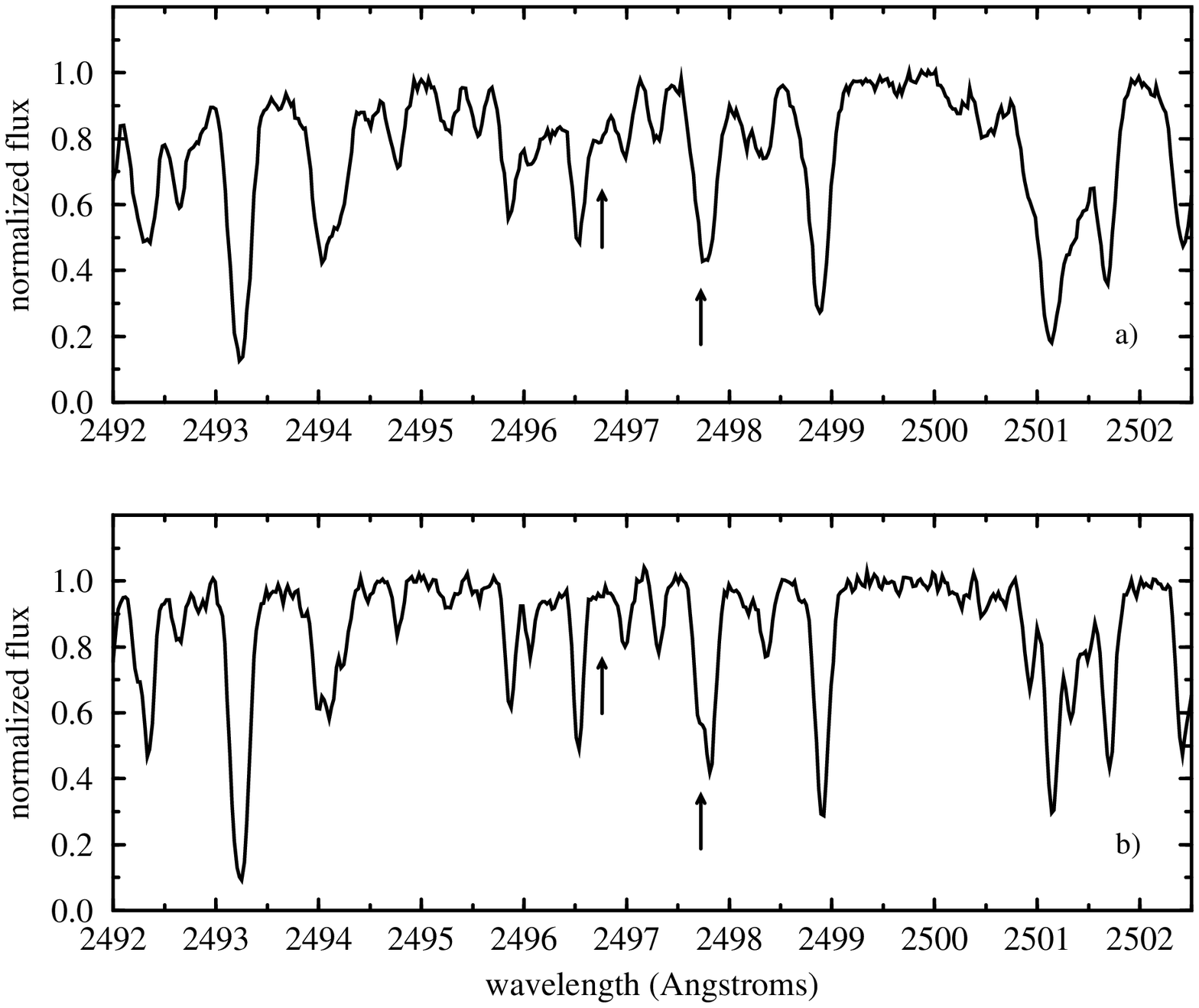}
\caption[b_fig1.eps]{The observed spectra of (a) BD $+23$\arcdeg 3130 and
(b) HD 84937 from 2492 to 2502.5 \AA. The locations of the \ion{B}{1} resonance
lines are indicated. \label{fig1}}
\end{figure}

\begin{figure}
\plotone{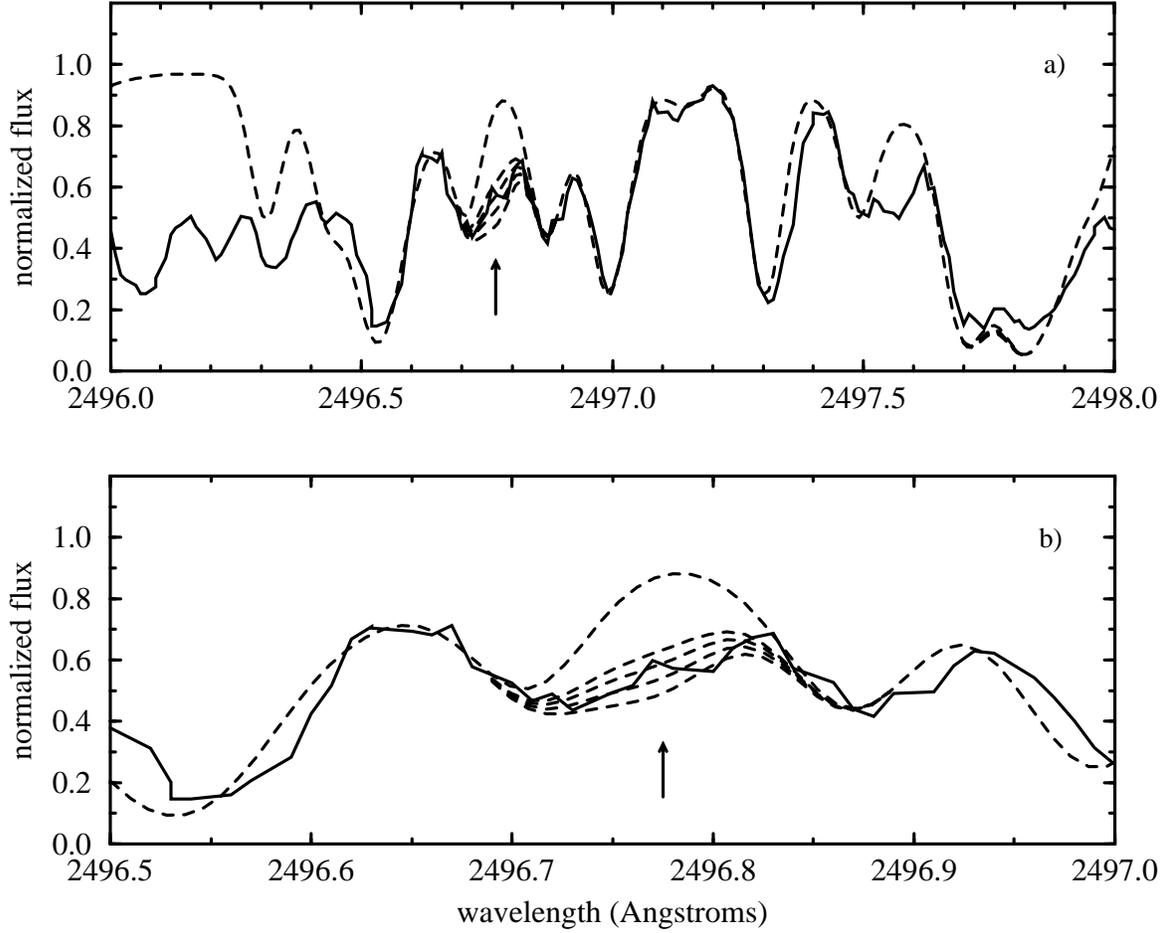}
\caption[b_fig2.eps]{(a) Comparison between the observed spectrum of Procyon
(solid line) and five synthetic spectra (dashed lines) computed with our line
list, using a Kurucz model photosphere, and log N(B)$=-3, 1.9, 2.0, 2.1$, and
2.2, respectively. (b) Zoom of the region close to the \ion{B}{1} $\lambda$
2496.772 \AA\ line showing the high sensitivity of the observed feature to
changes in the boron abundance of the synthetic profiles.
\label{fig2}}
\end{figure}

\begin{figure}
\plotone{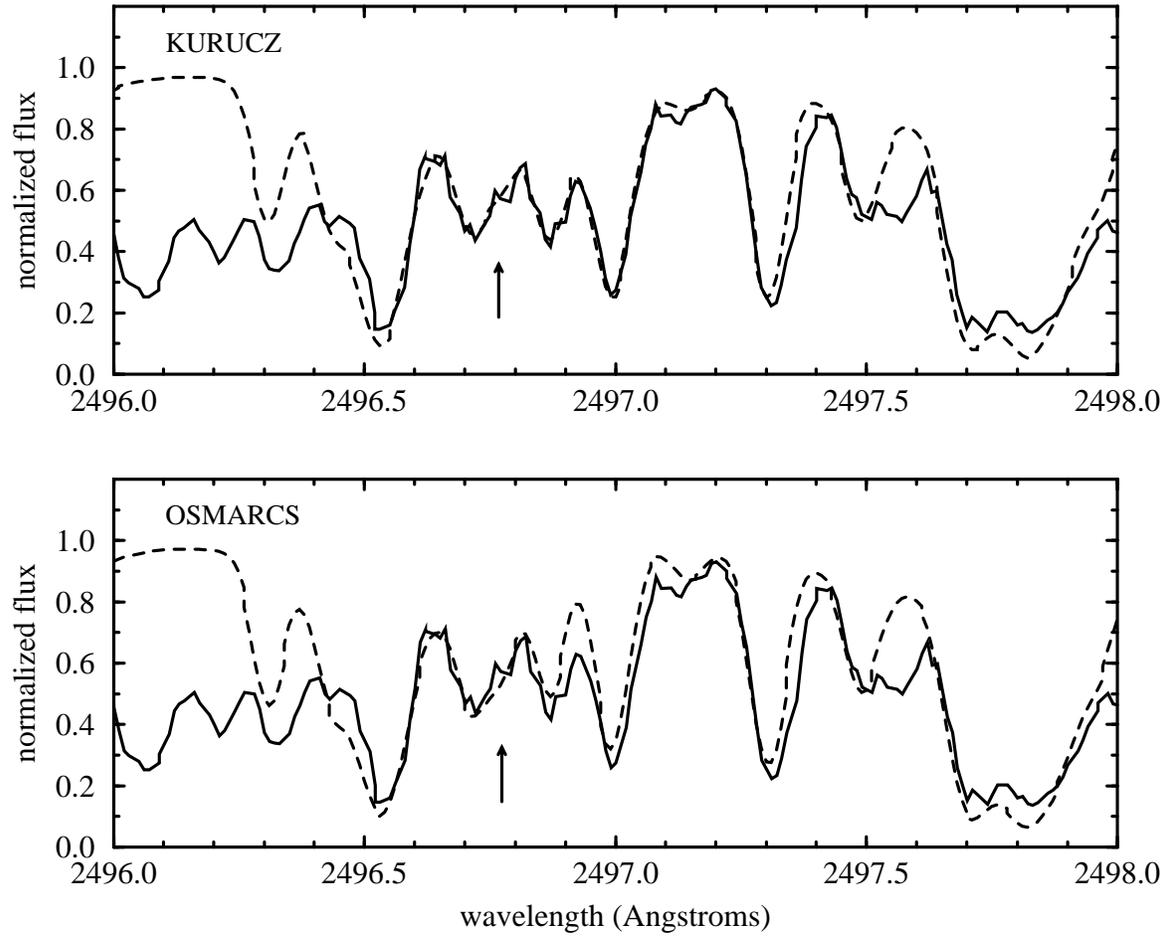}
\caption[b_fig3.eps]{Comparison between the observed spectrum of Procyon
(solid line) and two synthetic spectra (dashed lines) computed with the same
stellar parameters and a boron abundance log N(B) $=2.0$, using Kurucz and
OSMARCS model photospheres, respectively. \label{fig3}}
\end{figure}

\begin{figure}
\plotone{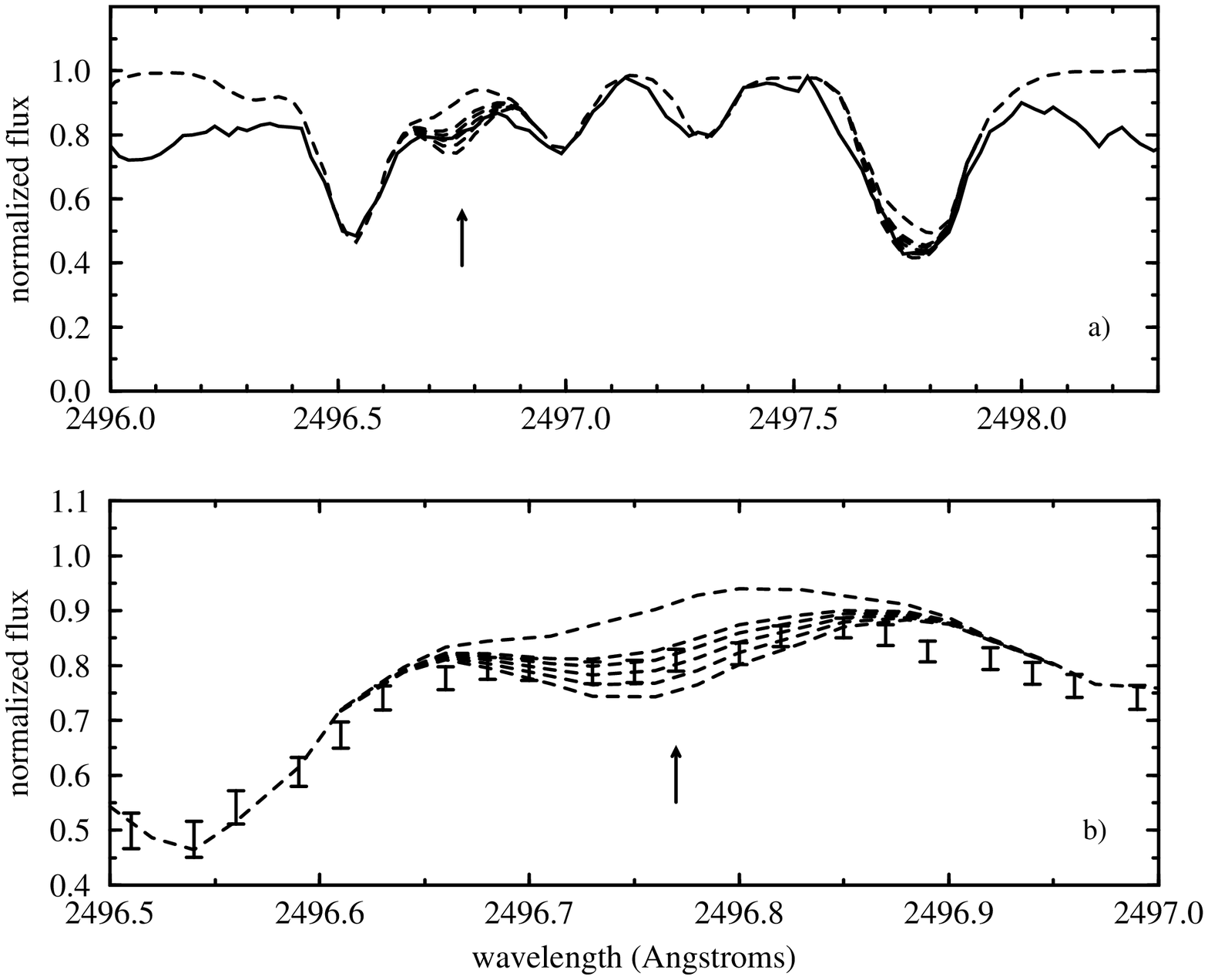}
\caption[b_fig4_2.eps]{(a) Comparison between the observed spectrum of BD
$+23$\arcdeg 3130  (solid line) and six synthetic spectra (dashed lines)
computed using a Kurucz model, with log N(B)$=-3, -0.5, -0.4, -0.3, -0.2$, and
$-0.1$, respectively. (b) Zoom of the region close to the \ion{B}{1} $\lambda$
2496.772 \AA\ line showing the high sensitivity of the observed feature to
changes in the boron abundance of the synthetic profiles (the observed spectrum
is represented by photon statistics error bars). \label{fig4}}
\end{figure}

\begin{figure}
\plotone{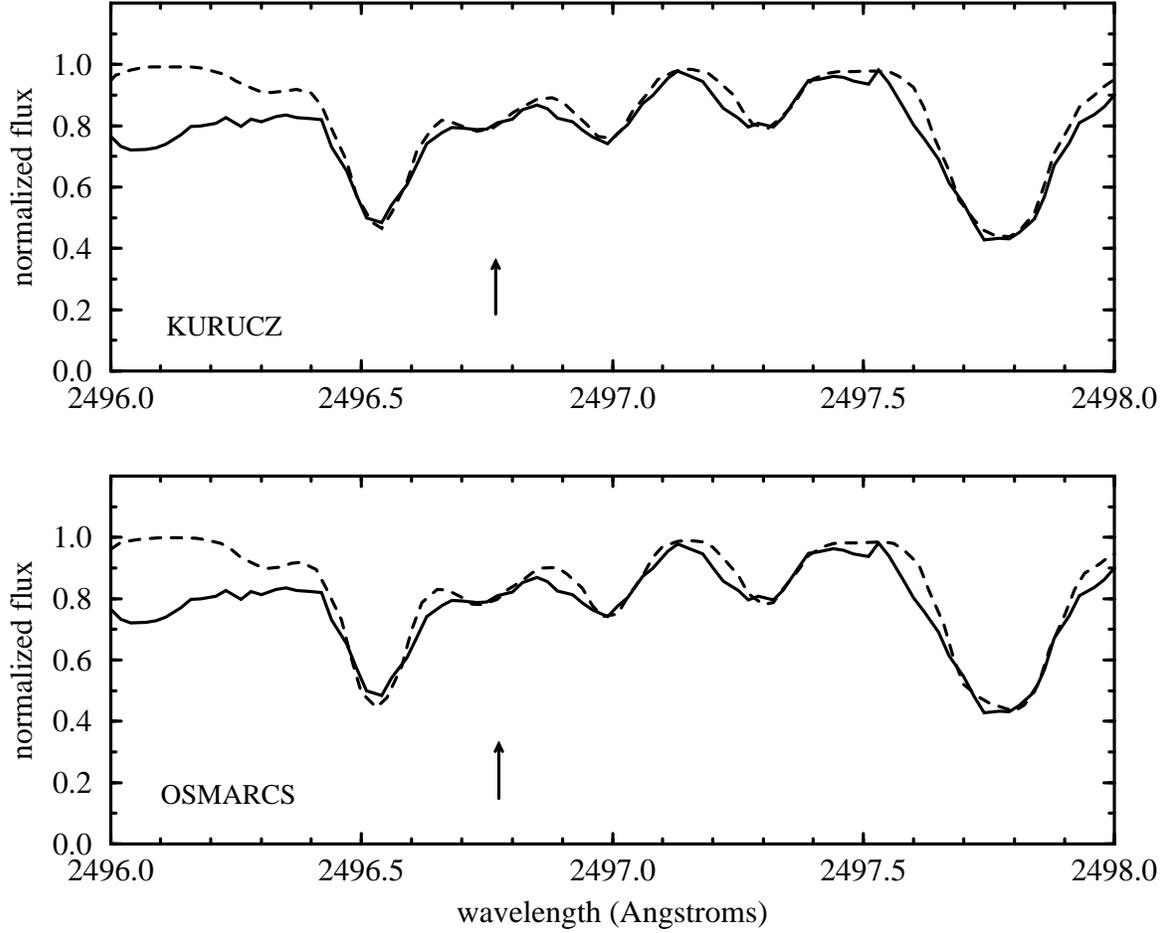}
\caption[b_fig5.eps]{Comparison between synthetic spectra (dashed
lines), computed using Kurucz (upper panel) and OSMARCS (lower panel) model
photospheres, which best matches the observed spectrum of BD $+23$\arcdeg 3130
(solid line). The synthetic spectrum in the lower panel was computed with
$-0.1$ dex in [Fe/H] and boron abundance with respect to the upper one. The
rest of stellar parameters used were the same in both cases. \label{fig5}}
\end{figure}

\begin{figure}
\plotone{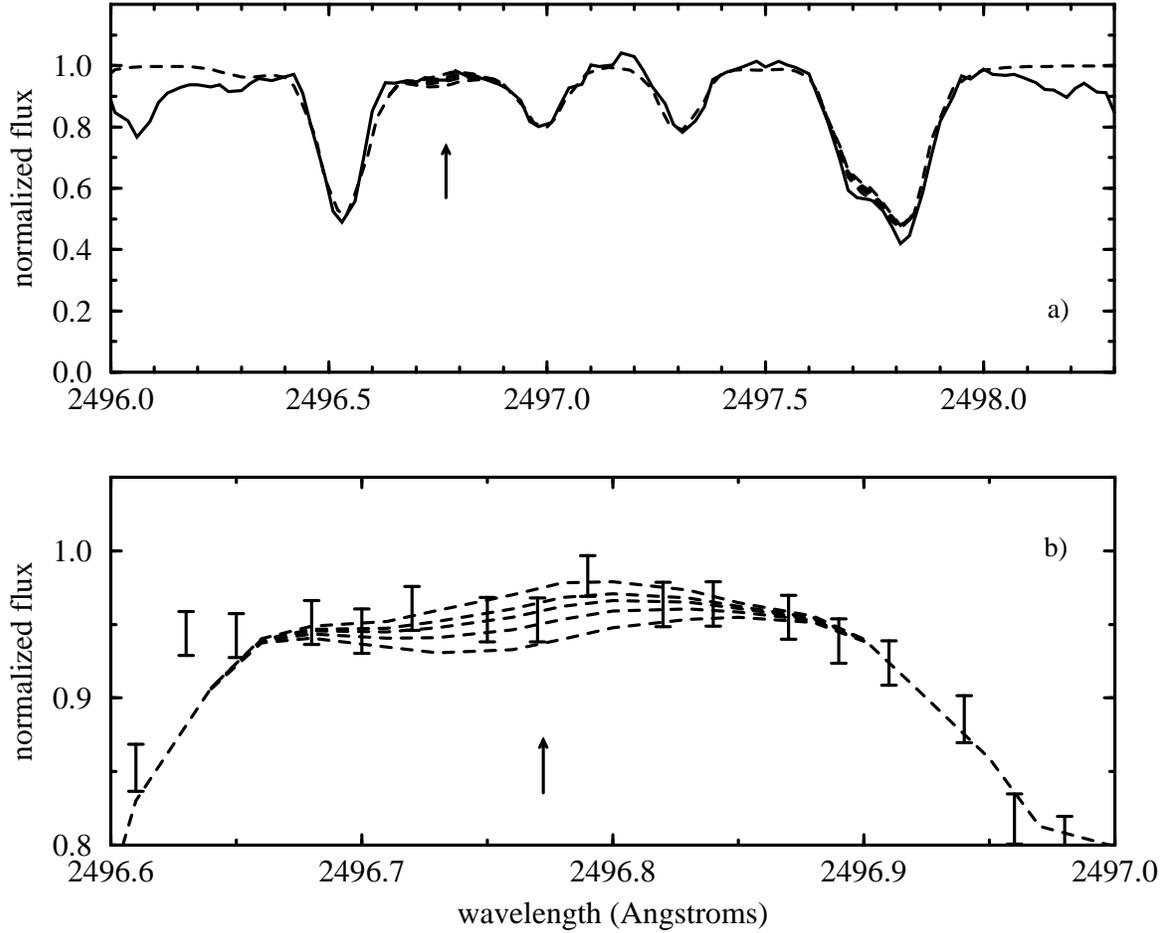}
\caption[b_fig6_2.eps]{(a) Comparison between the observed spectrum of HD 84937
(solid  line) and five synthetic spectra (dashed lines) computed using a Kurucz
model, with log  N(B)$=-3, -0.2, 0.0, 0.2$, and 0.4, respectively.  (b) Zoom of
the region close to the  \ion{B}{1} $\lambda$ 2496.772 \AA\ line showing the
low sensitivity of the  observed feature to changes in the boron abundance of
the synthetic profiles (the observed spectrum is represented by photon
statistics error bars). \label{fig6}}
\end{figure}

\begin{figure}
\plotone{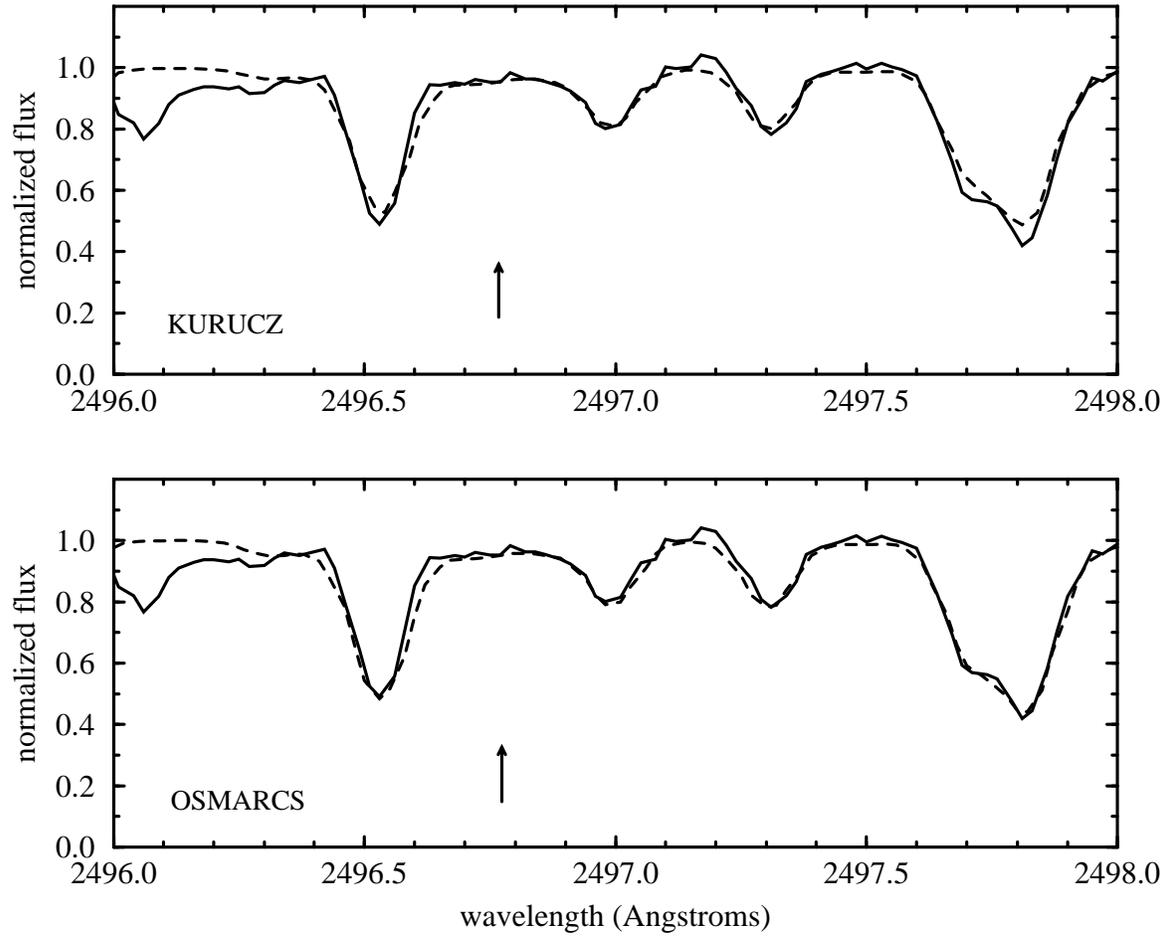}
\caption[b_fig7.eps]{Comparison between synthetic spectra (dashed
lines), computed using Kurucz (upper panel) and OSMARCS (lower panel) model
photospheres, which best matches the observed spectrum of HD 84937 (solid
line). Equal values of stellar parameters and boron abundance were used in both
cases. \label{fig7}}
\end{figure}

\begin{figure}
\plotone{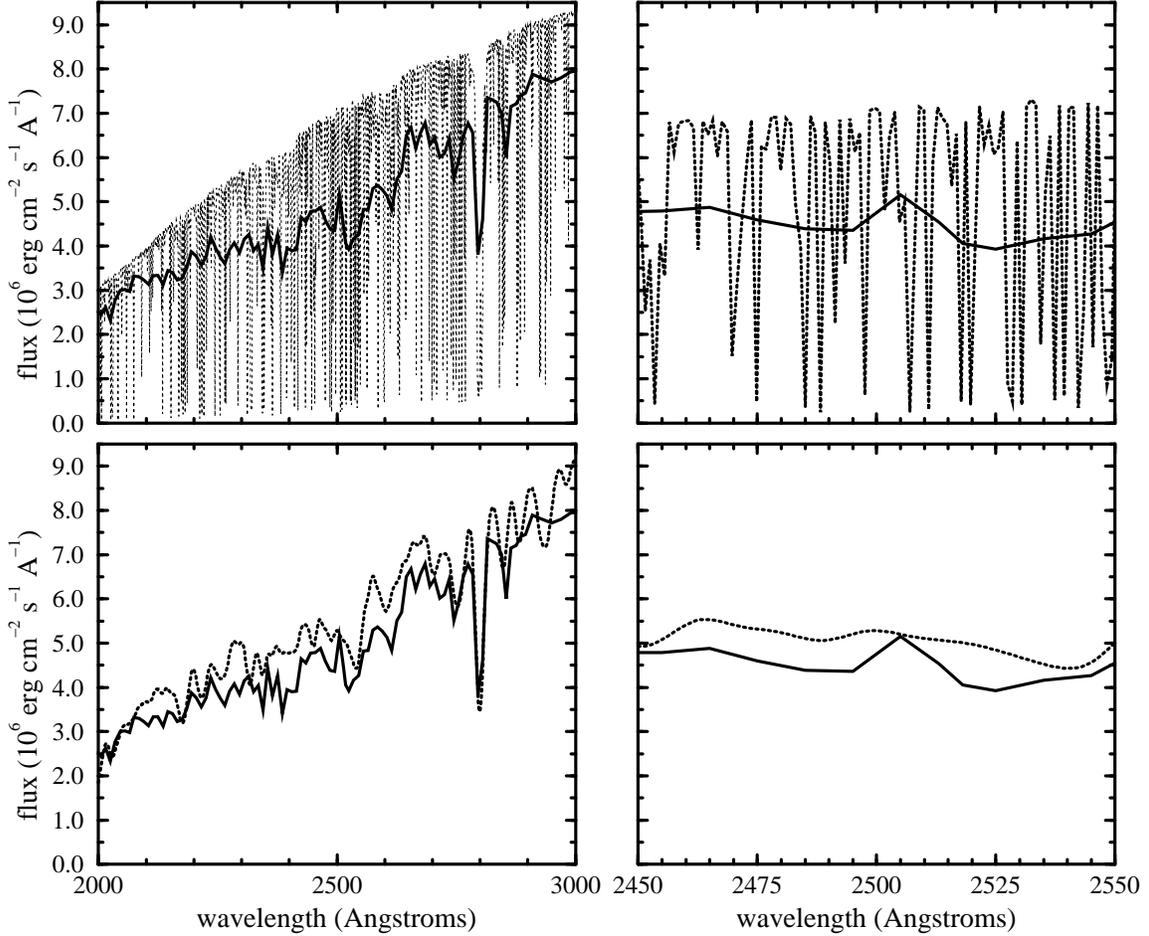}
\caption[b_fig8.eps]{Comparison between the energy fluxes predicted at
$\lambda$ 2500 \AA\ by Kurucz (solid lines) and OSMARCS (dotted lines) model
photospheres used in this work. Upper panels show a direct comparison in their
original form, indicating a significant difference of fluxes predicted by
models computed using opacity distribution functions and opacity sampling
techniques. In the lower panels the OSMARCS models have been degraded to match
the resolution of Kurucz ones, and the differences between models are reduced. 
\label{fig8}}
\end{figure}

\begin{figure}
\plotone{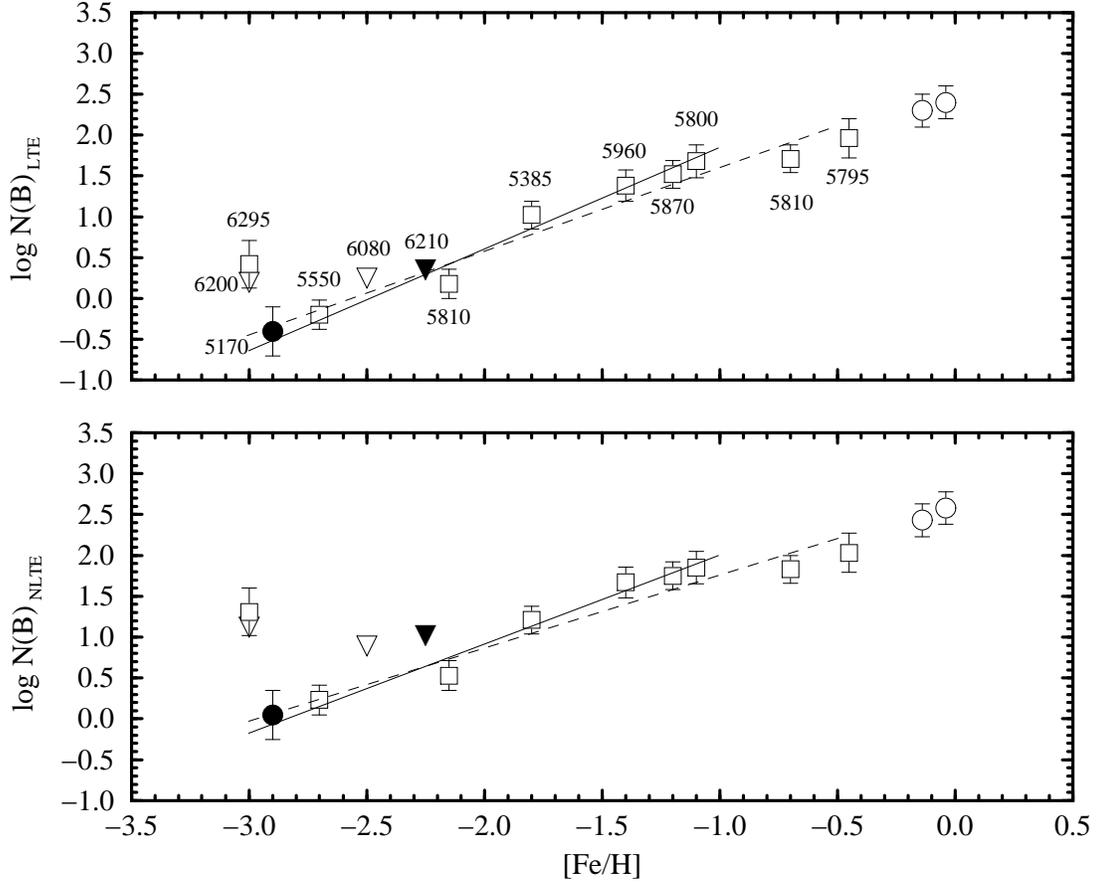}
\caption[b_fig9.eps]{LTE and NLTE boron abundances against metallicity
for all the stars  considered. Filled symbols, stars observed in this work;
open triangles,  BD $-13$\arcdeg 3442 (Rebull et al. 1996) and BD $+26$\arcdeg
3578 (Duncan et  al. 1997a) reanalyzed using spectral synthesis; open squares,
stars of Duncan  et al. (1997a) with abundances adapted to our stellar
parameters using  equivalent  widths. Inverted triangles represent upper limits
indicated by spectral  synthesis. The abundance of the star  BD $+3$\arcdeg 740
($T_{\rm eff}=6295$ K and [Fe/H]$=-3$) is also considered an upper limit (see
text). In the upper panel the stars are labeled with the  adopted effective
temperatures. Also included for comparison are the boron abundances of the
solar-metallicity stars $\iota$ Peg and $\theta$ UMa taken from Lemke et al.
(1993; open circles). Dashed lines correspond to linear least squares  fits to
those metal-poor stars with $T_{\rm eff}<6000$ K, and solid lines  restrict the
fits to stars with $T_{\rm eff}<6000$ K and [Fe/H]$<-1$. \label{fig9}}
\end{figure}

\begin{figure}
\plotone{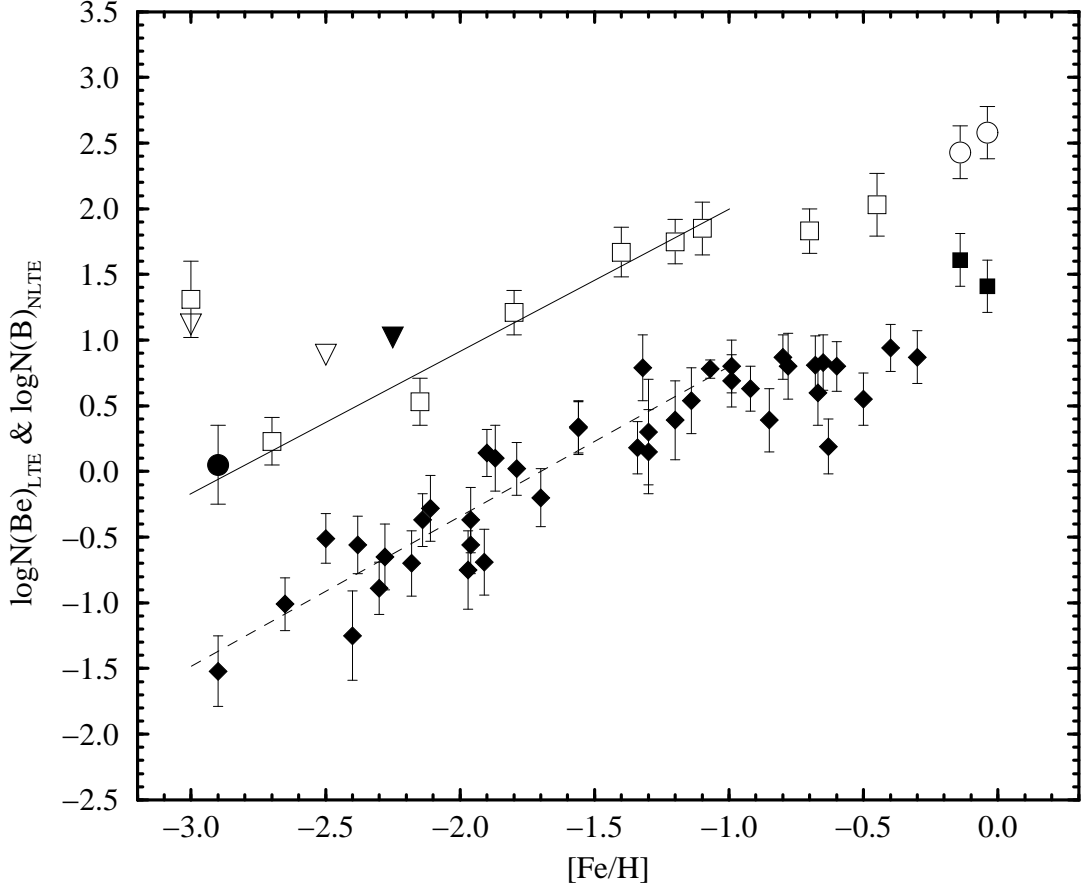}
\caption[b_fig10.eps]{Boron (NLTE) and beryllium abundances against
metallicity.  Beryllium data were taken from Garc\'\i a L\'opez et al. (1998),
and represent a large sample of metal-poor stars with abundances derived using 
high-resolution CCD spectra. Be abundances for the solar-metallicity stars
$\iota$ Peg and $\theta$ UMa, taken from Lemke et al. (1993; filled squares),
are shown for comparison. The solid line is a least squares linear fit to boron
abundances of stars with $T_{\rm eff}<6000$ K and [Fe/H]$<-1$, and the dashed
line correspond to a similar fit to the beryllium data. \label{fig10}}
\end{figure}

\begin{figure}
\plotone{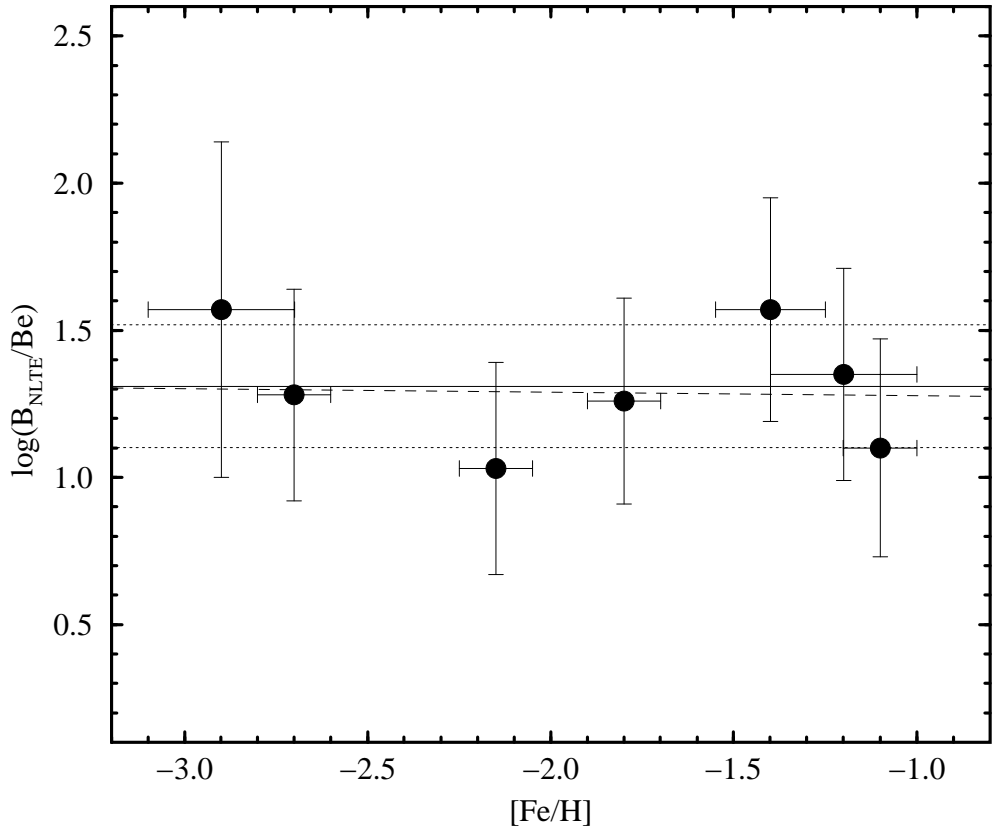}
\caption[b_fig11.eps]{Boron (NLTE) to beryllium (LTE) ratios against
metallicity for those seven stars with $T_{\rm eff}<6000$ K and [Fe/H]$<-1$.
\label{fig11}}
\end{figure}

\clearpage

\begin{deluxetable}{lccccccc}
\tablecaption{Stellar Parameters and Boron Abundances \label{table1}}
\tablewidth{0pt}
\small
\tablehead{
\colhead{Star} & \colhead{$T_{\rm eff}$} & \colhead{$\log g$} &\colhead{[Fe/H]} 
& \colhead{log N(B)} & \colhead{log N(B)} & \colhead{log N(B)}
 & \colhead{Ref.} \nl
 & & & & \colhead{LTE} & \colhead{LTE} & \colhead{NLTE} & \nl
 & & & & KURUCZ & OSMARCS & OSMARCS & }
\startdata
BD $+23$\arcdeg 3130 & 5170$\pm$150 & 2.50$\pm$0.30 & $-2.90\pm 0.20$ & 
$-0.30\pm 0.30$     & $-0.40$   & \phm{$<$}0.05 & 1,a  \nl
HD 84937             & 6210$\pm$120 & 4.00$\pm$0.10 & $-2.25\pm 0.15$ & 
$<0.35$             & $<0.35$   & $<1.02$	& 1,a  \nl
BD $-13$\arcdeg 3442 & 6200$\pm$150 & 3.50$\pm$0.20 & $-3.00\pm 0.20$ & 
$<0.20$             & $<0.20$   & $<1.12$	& 2,a  \nl
BD $+26$\arcdeg 3578 & 6080$\pm$120 & 3.85$\pm$0.10 & $-2.50\pm 0.10$ & 
$<0.25$             & $<0.25$   & $<0.89$	& 3,a  \nl
HD 142373            & 5795$\pm$100 & 4.30$\pm$0.10 & $-0.45\pm 0.10$ & 
\phs $1.99\pm 0.24$ & \phs 1.96 & \phm{$<$}2.03 & 3,b  \nl
HD 184499            & 5810$\pm$100 & 4.00$\pm$0.10 & $-0.70\pm 0.10$ & 
\phs $1.70\pm 0.17$ & \phs 1.71 & \phm{$<$}1.83 & 3,b  \nl
HD 76932             & 5800$\pm$100 & 3.85$\pm$0.10 & $-1.10\pm 0.10$ & 
\phs $1.77\pm 0.20$ & \phs 1.68 & \phm{$<$}1.85 & 3,b  \nl
HD 201891            & 5870$\pm$100 & 4.00$\pm$0.10 & $-1.20\pm 0.20$ & 
\phs $1.62\pm 0.17$ & \phs 1.52 & \phm{$<$}1.75 & 4,b  \nl
HD 194598            & 5960$\pm$120 & 4.15$\pm$0.10 & $-1.40\pm 0.15$ & 
\phs $1.42\pm 0.19$ & \phs 1.38 & \phm{$<$}1.67 & 3,b  \nl
HD 64090             & 5385$\pm$100 & 4.45$\pm$0.10 & $-1.80\pm 0.10$ & 
\phs $1.07\pm 0.17$ & \phs 1.02 & \phm{$<$}1.21 & 3,b  \nl
HD 19445             & 5810$\pm$150 & 4.30$\pm$0.10 & $-2.15\pm 0.10$ & 
\phs $0.21\pm 0.18$ & \phs 0.18 & \phm{$<$}0.53 & 4,b  \nl
HD 140283            & 5550$\pm$100 & 3.35$\pm$0.10 & $-2.70\pm 0.10$ & 
$-0.15\pm 0.18$     & $-0.20$   & \phm{$<$}0.23 & 4,b  \nl
BD $+3$\arcdeg 740   & 6295$\pm$120 & 4.00$\pm$0.10 & $-3.00\pm 0.10$ & 
\phs $0.47\pm 0.29$ & \phs 0.42 & \phm{$<$}1.31 & 3,b  \nl
\enddata
\tablerefs{1, this work; 2, star observed by Rebull et al. (1996); 3, by Duncan
et al. (1997a); 4, by Duncan, Lambert, \& Lemke (1992).}
\tablerefs{a, LTE boron abundance derived by spectral synthesis; b, using
estimated equivalent widths.}

\end{deluxetable}

\clearpage

\begin{deluxetable}{lccc}
\tablecaption{Beryllium and Lithium Abundances for those Stars with 
$T_{\rm eff}<6000$ K and [Fe/H]$<-1$ \label{table2}}
\tablehead{
\colhead{Star} & \colhead{[Fe/H]} & \colhead{log N(Be)} & \colhead{log N(Li)} 
\nl
 & & \colhead{(LTE)} & \colhead{(NLTE)} } 
\startdata
HD 76932             & $-1.10\pm 0.10$ & \phs $0.75\pm 0.17$ & $2.02\pm 0.10$ 
\nl
HD 201891            & $-1.20\pm 0.20$ & \phs $0.40\pm 0.19$ & $1.98\pm 0.10$ 
\nl
HD 194598            & $-1.40\pm 0.15$ & \phs $0.10\pm 0.19$ & $2.14\pm 0.10$ 
\nl
HD 64090             & $-1.80\pm 0.10$ & $-0.05\pm 0.18$     & $1.34\pm 0.08$ 
\nl
HD 19445             & $-2.15\pm 0.10$ & $-0.50\pm 0.18$     & $2.12\pm 0.13$ 
\nl
HD 140283            & $-2.70\pm 0.10$ & $-1.05\pm 0.18$     & $2.14\pm 0.10$ 
\nl
BD $+23$\arcdeg 3130 & $-2.90\pm 0.20$ & $-1.52\pm 0.27$     & $1.30\pm 0.10$ 
\nl
\enddata
\end{deluxetable}


\begin{references}

\reference{Abia} Abia, C., \& Rebolo, R. 1989, \apj, 347, 186

\reference{Alonso} Alonso, A., Arribas, S., \& Mart\'\i nez-Roger, C. 1996a, 
\aap, 313, 873

\reference{Alonso} Alonso, A., Arribas, S., \& Mart\'\i nez-Roger, C. 1996b,
\aaps, 117, 227

\reference{Bessell} Bessell, M. S., Sutherland, R. S., \& Ruan, K. 1991, \apjl,
L71

\reference{Blackwell} Blackwell, D. E., Petford, A. D., Arribas, S., Haddock, 
D. J., \& Selby, M. J. 1990, \aap, 232, 396

\reference{Boesgaard} Boesgaard, A. M. 1976, \apj, 210, 466

\reference{BH} Boesgaard, A. M., \& Heacox, W. D. 1978, \apj, 226, 888

\reference{Carlsson} Carlsson. M., Rutten, R. J., Bruls, J. H. M. J., \&
Shchukina, N. G. 1994, \aap, 288, 860

\reference{C83} Carney, B. W. 1983, \aj, 88, 623

\reference{Carney} Carney, B. W., Latham, D. W., Laird, J. B, \& Aguilar, L.
A. 1994, \aj, 107, 2240

\reference{CB} Casuso, E., \& Beckman, J. E. 1997, \apj, 475, 155

\reference{Cavallo} Cavallo, R. M., Pilachowski, C. A., \& Rebolo, R. 1997, 
\pasp, 109, 226

\reference{Chaussidon} Chaussidon, M., \& Robert, F. 1995, \nat, 374, 337

\reference{DLL} Duncan, D. K., Lambert, D. L., \& Lemke, M. 1992, \apj, 401, 
584

\reference{Duncan97b} Duncan D. K., Peterson, R. C., Thorburn, J. A., 
Pinsonneault, M. H., \& Deliyannis, C. P. 1997b, \apj\ (in press)

\reference{Duncan97a} Duncan, D. K., Primas, F., Rebull, L. M., Boesgaard, 
A. M., Deliyannis, C. P., Hobbs, L. M., King, J. R., \& Ryan, S. G. 1997a, 
\apj, 488, 338

\reference{Duncan98} Duncan, D. K., Rebull, L. M., Primas, F., Boesgaard, 
A. M., Deliyannis, C. P., Hobbs, L. M., King, J. R., \& Ryan, S. G. 1998,
\aap\ (in press)

\reference{Edvardsson93} Edvardsson, B., Andersen, J., Gustafsson, B.,
Lambert, D. L., Nissen, P. E., \& Tomkin, J. 1993, \aap, 275, 101
 
\reference{Edvardsson} Edvardsson, B. Gustafsson, B., Johansson, S. G., 
Kiselman, D., Lambert, D. L., Nissen, P. E., \& Gilmore, G. 1994, \aap, 290, 176

\reference{Feltzing} Feltzing, S., \& Gustafsson, B. 1994, \apj, 423, 68

\reference{GarciaLopez} Garc\'\i a L\'opez, R. J., Rebolo, R., P\'erez de Taoro,
 M. R., \& Alonso, A. 1998, \aap\ (in preparation)
 
\reference{BeNLTE} Garc\'\i a L\'opez, R. J., Severino, G., \& Gomez, M. T.
1995, \aap, 297, 787  

\reference{Gilmore} Gilmore, G., Gustafsson, B., Edvardsson, B.,\&
 Nissen, P. E. 1992, Nature, 357, 392 
 
\reference{Gratton} Gratton, R., Caretta, E., \& Castelli, F. 1996, \aap, 314,
 191 

\reference{Grevesse} Grevesse, N., Noels, A., \& Sauval, A. J. 1996,
   ASP Conf. Ser., 99, 117
   
\reference{Gustafsson} Gustafsson, B., Bell, R. A., Eriksson, K., \& Nordlund,
\AA\ 1975, \aap, 42, 407
   
\reference{HT94} Hobbs, L. M., \& Thorburn, J. A. 1994, \apjl, 428, L25

\reference{HT97} Hobbs, L. M., \& Thorburn, J. A. 1997, \apj\ (in press) 

\reference{Israelian} Israelian, G., Garc\'\i a L\'opez, R. J., \& Rebolo, R.
1998, \apj\ (in preparation)
    
\reference{Jedamzik} Jedamzik, K., Fuller, G. M., Matheus, G. J., \& Kajino, T.
1994, \apj, 422, 423

\reference{Johansson} Johansson, S. G. 1992, private communication

\reference{JC} Johansson, S. G., \& Cowley 1988, J. Opt. Soc. Am. B, 5, 2264

\reference{Johansson et al.} Johansson, S. G., Litz\'en, U., Kasten, J., \&
Kock, M. 1993, \apjl, 403, L25 

\reference{Kajino} Kajino, T., \& Boyd, R. N. 1990, \apj, 359, 267

\reference{Kiselman} Kiselman, D. 1994, \aap, 286, 159

\reference{KC95} Kiselman, D., \& Carlsson, M. 1995, in The Light Element
Abundances, ed. P. Crane (Springer), 372

\reference{KC} Kiselman, D., \& Carlsson, M. 1996, \aap, 311, 680

\reference{Kurucz92} Kurucz, R. L. 1992, private communication

\reference{Kurucz93} Kurucz, R. L. 1993, CD-ROM \# 1

\reference{Laird} Laird, J. B., Carney, B., \& Latham, D. W. 1988, \aj, 95,
1843

\reference{Lemke} Lemke, M., Lambert, D. L., \& Edvardsson, B. 1993, \pasp, 
105, 468

\reference{Malaney} Malaney, R. A., \& Fowler, W. A. 1989, \apjl, 345, L5

\reference{Molaro} Molaro, P. 1987, \aap, 183, 241

\reference{Molaro97} Molaro, P., Bonifacio, P., Castelli, F., \& Pasquini, L.
1997, \aap, 319, 593

\reference{Nissen94} Nissen, P. E., Gustafsson, b., Edvardsson, b., \& Gilmore,
G. 1994, \aap, 285, 400

\reference{Nissen97} Nissen, P. E., H$\o$g, E., \& Schuster, W. J. 1997,
proceedings of the ``Hipparcos Venice'97 Symposium'', ESA SP-402 (in press)

\reference{O'Brian} O'Brian, T. R., \& Lawler, J. E. 1992, \aap, 255, 420

\reference{O'BFe} O'Brian, T. R., Wickliffe, M. E., Lawler, J. G.,
  Whaling, W., Brault, J. W. 1991, J. Opt. Soc. Am. B8, 1185

\reference{Pavlenko} Pavlenko, Ya. V. 1991, SvA, 35, 212

\reference{Pavlenko96} Pavlenko, Ya. V., \& Magazz\`u, A. 1996, \aap, 311, 961

\reference{Pavlenko95} Pavlenko, Ya. V., Rebolo, R., Mart\'\i n, E. L., \&
Garc\'\i a L\'opez, R. J. 1995, \aap, 303, 807

\reference{Press} Press, W. H., Teukolsky, A. A., Vetterling, W. T., \&
Flannery, B. P. 1992, Numerical Recipes, 2nd. ed., Cambridge University Press,
Cambridge

\reference{Ramaty96} Ramaty, R., Kozlovsky, B., \& Lingenfelter, R. E. 1996,
 \apj, 456, 525
 
\reference{Ramaty97} Ramaty, R., Kozlovsky, B., Lingenfelter, R. E., \&
Reeves, H. 1997, \apj, 488, 730

\reference{Rebull} Rebull, L. M., Duncan, D. K., Boesgaard, A. M., Deliyannis, 
C. P., Hobbs, L. M., King, J. R., \& Ryan, S. G. 1996, ASP Conf. Ser. 99, 184

\reference{Reeves} Reeves, H. 1994, Rev. Mod. Phys. 66, 193

\reference{Reevesetal} Reeves, H., Richer, J., Terasawa, N., \& Sato, K. 1990,
\apj, 355, 18

\reference{Ryan} Ryan, S. G., Beers, T. C., Deliyannis, C. P., \& Thorburn, J.
A. 1996, \apj, 458, 543

\reference{SLN} Smith, V., Lambert, D. L., \& Nissen, P. E. 1993, \apj, 408, 262

\reference{Stryker} Stryker, L. L., Hesser, J. E., Hill, G., Garlick, G. S.,
\& O'Keefe, L. M. 1985, \pasp, 97, 247

\reference{Terasawa} Terasawa, N., \& Sato, K. 1990, \apjl, 362, L47

\reference{Thomas93} Thomas, D., Schramm, D. N., Olive, K. A., \& Fields,
B. D. 1993, \apj, 406, 569

\reference{Thomas94} Thomas, D., Schramm, D. N., Olive, K. A., Mathews, G. J.,
 Meyer, B. S., \& Fields, B. D. 1994, \apj, 430, 291

\reference{Thorburn} Thorburn, J. A. 1994, \apj, 421, 318

\reference{Timmes} Timmes, F. X., Woosley, S. E., \& Weaver, T. A. 1995,
 \apjs, 98, 617

\reference{VF} Vangioni-Flam, E., Cass\'e, M., Fields, B. D., \& Olive, K. A. 
1996, \apj, 468, 199 

\end{references}
\end{document}